\documentclass[10pt]{article}
\usepackage{aimsreport,times}

\usepackage{amsmath,amsfonts,bm}

\def\eqref#1{equation~\ref{#1}}

\def\1{\bm{1}}

\DeclareMathAlphabet{\mathsfit}{\encodingdefault}{\sfdefault}{m}{sl}
\SetMathAlphabet{\mathsfit}{bold}{\encodingdefault}{\sfdefault}{bx}{n}

\newcommand{\iclrvspace}[1]{}
\newcommand{\iclrarxiv}[2]{#2}

\usepackage{hyperref}
\usepackage{url}
\usepackage{booktabs}
\usepackage{array}
\usepackage{amssymb}
\usepackage{wrapfig}
\usepackage{makecell}
\usepackage{threeparttable}
\usepackage{soul}

\usepackage{subcaption}
\usepackage{tabularx}
\usepackage{pifont}
\usepackage{multirow}
\usepackage{listings}
\usepackage{enumitem}
\newtcolorbox{promptbox}[2][]{breakable, fonttitle=\bfseries\footnotesize, #1, #2,
  colback=aimsbg, colframe=aimsaccent, boxrule=0.6pt, arc=2pt}

\title{VoxMem: Benchmarking Multimodal Memory\\in Large Audio Language Models}
\shorttitle{VoxMem: Benchmarking Multimodal Memory in Large Audio Language Models}
\author{Yang~Xiao\textsuperscript{1} \and Vidhyasaharan~Sethu\textsuperscript{2} \and
  Eun-Jung~Holden\textsuperscript{1} \and Ting~Dang\textsuperscript{1}}
\affiliations{\textsuperscript{1}University of Melbourne\qquad
  \textsuperscript{2}University of New South Wales}
\reportlab{AIMS Lab}
\reportdate{September 2026}
\reportlink[\faGithub]{Code}{https://github.com/swagshaw/voxmem/tree/main}{github.com/swagshaw/voxmem}
\reportlink[\faDatabase]{Data}{https://huggingface.co/datasets/AudioMemory/voxmembench}{huggingface.co/datasets/AudioMemory/voxmembench}

\hypersetup{pdftitle={VoxMem: Benchmarking Multimodal Memory in Large Audio Language Models},
  pdfauthor={Yang Xiao, Vidhyasaharan Sethu, Eun-Jung Holden, Ting Dang}}

\begin{document}

\maketitle
\begin{abstract}
Spoken conversational systems must recover information from prior interactions (i.e., memory), yet relevant information in speech extends beyond what was said to who said it, how it was spoken, and what was audible, information that exists only in the audio signal and cannot be recovered from a transcript. Beyond what to remember, memory also demands diverse operations: retrieving a single fact, integrating evidence across turns, tracking an evolving state. Real interactions further unfold across sessions, meaning information accumulates across distinct episodes rather than a single continuous recording. Existing benchmarks fall short on all three dimensions: they focus primarily on lexical content, adopt limited and ad hoc memory operations, and treat memory as a single-session problem. We argue that principled memory evaluation requires jointly characterizing the acoustic evidence to be retained and the operations applied to it, and introduce a taxonomy along these two axes. Building on this taxonomy, we present VoxMem: 3,196 evaluation instances over 34,743 spoken sessions (177 hours)
crossing four acoustic evidence types (speech semantics, speaker identity, paralinguistic cues, environmental sound) with four memory operations (information extraction, multi-session reasoning, temporal tracking, and answer refusal), grounded in multi-session histories and stratified across context budgets from 8K to 64K tokens. Evaluating 15 LALMs, no model exceeds 40\% at 32K. Models retain \textit{what} was said far better than \textit{who} said it, \textit{how}, or \textit{what} was audible, a gap that widens for complex operations, grows with history length, and manifests as qualitatively distinct failure modes across evidence types. VoxMem aims to provide a foundation to measure and drive progress on the full scope of spoken conversational memory.
\end{abstract}

\section{Introduction}
Large audio language models (LALMs) have rapidly advanced natural spoken interaction, increasingly enabling dialogue systems that span extended, multi-session conversations~\citep{chu2023qwenaudio,tang2024salmonn,ghosh2025audioflamingo2,wu2025stepaudio2}. A fundamental requirement for such systems is long-term memory: the ability to accumulate information across past interactions and ground present responses in that history. One approach is long-context scaling: expanding the context window so that longer dialogue histories can be processed directly~\citep{kim2025contextdialog,lin2025speechprune,he2026headrouter,ghosh2026audioflamingonext}.
While conceptually straightforward, providing a longer history does not guarantee that the model can reliably retrieve and use the information it contains.
 As interaction histories grow, whether information is accurately retrieved and used across sessions becomes essential for evaluating spoken conversational memory and guiding future memory systems.


\textbf{Existing benchmarks lack a principled taxonomy of audio memory and cover only a narrow range of spoken-memory capabilities.} As summarized in Table~\ref{tab:benchmark_comparison_full}, prior benchmarks assess linguistic content and selected acoustic cues, but rarely test whether models can remember \emph{who} spoke, \emph{how} they spoke, or \emph{what} was audible in the environment. Along a second dimension, they provide only limited and fragmented coverage of memory operations, including retrieval, evidence integration, temporal tracking, and recognizing insufficient evidence~\citep{wu2025longmemevalbenchmarkingchatassistants,ren2025memlens,wu2026longmemeval}. Consequently, the acoustic evidence and memory operations assessed by existing benchmarks are selected in an ad hoc manner, rather than derived from a unified and principled account of spoken memory. Without a taxonomy that jointly characterizes the acoustic evidence to be remembered and the memory operations, acoustic memory cannot be evaluated systematically.


\textbf{Cross-session memory, the ability to retain and relate information across distinct conversational episodes, remains insufficiently characterized in existing spoken-language evaluations.} 
As shown in Table~\ref{tab:benchmark_comparison_full}, existing benchmarks primarily evaluate isolated recordings, and recently have begun to consider continuous dialogues. Yet real human--machine spoken interactions unfold across separate sessions, with changing topics, contexts, and temporal gaps~\citep{maharana2024evaluatinglongtermconversationalmemory,kim2025contextdialog}. This setting requires models to retrieve and relate evidence dispersed across a long, heterogeneous history, for example, linking recurring speakers to prior statements and tracking updates across sessions~\citep{jang2023conversation,maharana2024evaluatinglongtermconversationalmemory,wu2026longmemeval}. Whether LALMs can sustain such persistent recall as interaction histories grow remains open. Moreover, benchmarks that analyze history length typically compare different questions at different lengths~\citep{he2025audiomarathon,cheng2026voxinfinity}, confounding history length with question difficulty and evidence type. 

\newcommand{\cmark}{\ding{51}}
\newcommand{\xmark}{\ding{55}}
\newcommand{\partialmark}{\ensuremath{\odot}}
\newcommand{\bench}{VoxMem}

\definecolor{softred}{RGB}{200,60,60}
\definecolor{softredbg}{RGB}{253,232,232}

\begin{table*}[t]
\centering
\caption{\textbf{Comparison of spoken-memory benchmarks across evidence grounding, memory evaluation, controlled scaling, and history structure.} \textit{Answer-critical acoustic evidence} indicates whether answering a question requires different acoustic information. IE, MSR, TET, and AR denote the memory operations defined in Section 3.1. \textit{Prov.} denotes evidence provenance; \textit{Mono.} and \textit{Dial.} denote monologue and dialogue, respectively. }
\iclrvspace{-3mm}
\label{tab:benchmark_comparison_full}

\vspace{1mm}

{\setlength{\tabcolsep}{4.2pt}
\resizebox{\textwidth}{!}{%
\small
\begin{tabular}{
    @{}
    l
    c
    c
    cc
    cc
    @{}
}
\toprule

&
\textbf{Evidence Grounding}
&
\textbf{Memory Evaluation}
&
\multicolumn{2}{c}{\textbf{Controlled Scaling}}
&
\multicolumn{2}{c}{\textbf{History Structure}}
\\

\cmidrule(lr){2-2}
\cmidrule(lr){3-3}
\cmidrule(lr){4-5}
\cmidrule(lr){6-7}

\textbf{Benchmark}
&
\makecell{\textbf{Answer-critical}\\\textbf{Acoustic Evidence}}
&
\makecell{\textbf{Memory Operations}\\\textbf{IE/MSR/TET/AR}}
&
\textbf{Prov.}
&
\makecell{\textbf{Reported}\\\textbf{Length}}
&
\textbf{Type}
&
\makecell{\textbf{Multi-}\\\textbf{Sess.}}
\\

\midrule

AudioMarathon~\citep{he2025audiomarathon}
& \partialmark & IE & \xmark & 90--300\,s & Mono. & \xmark \\

LongSpeech~\citep{yang2026longspeech}
& \partialmark & IE & \xmark & $\sim$10\,min & Mono. & \xmark \\


VoiceGiraffe~\citep{ye2026voicegiraffe}
& \partialmark & IE/MSR & \xmark & 55.2\,min avg. & Mono. & \xmark \\

\addlinespace[1pt]

SpokenWOZ~\citep{si2023spokenwoz}
& \xmark & IE/MSR & \xmark & --- & Dial & \xmark \\

ContextDialog~\citep{kim2025contextdialog}
& \xmark & IE & \cmark & --- & Dial & \xmark \\

MTalk-Bench~\citep{du2025mtalkbench}
& \partialmark & IE/MSR & \xmark & --- & Dial & \xmark \\

Audio MultiChallenge~\citep{gosai2026audiomultichallenge}
& \partialmark & MSR/TET & \xmark & $8+$\,min & Dial & \xmark \\

Vox-Infinity~\citep{cheng2026voxinfinity}
& \partialmark & IE/MSR & \cmark & $\sim$23\,min & Dial/Mono & \xmark \\

\midrule

\textbf{\bench{} (Ours)}
& \textbf{\cmark} & \textbf{IE/MSR/TET/AR} & \textbf{\cmark} & \textbf{2.5--20\,min} & \textbf{Dial} & \textbf{\cmark} \\

\bottomrule
\end{tabular}%
}}
\iclrvspace{-8mm}
\end{table*}

To address these gaps, we introduce VoxMem, a benchmark for evaluating spoken conversational memory across two complementary axes: \textit{acoustic evidence}, specifying what must be recovered from the interaction history, and \textit{memory operation}, specifying how that information must be used.
Specifically, VoxMem spans four memory operations: information extraction~\citep{wu2025longmemevalbenchmarkingchatassistants,ren2025memlens}, multi-session reasoning, temporal evolution tracking, and answer refusal, and four information types: speech semantics, speaker identity, paralinguistic cues, and environmental sound. 
Established under the multi-session setting, we curate VoxMem with extensive quality control, comprising 3,196 evaluation instances over 34,743 spoken sessions (177 hours). To enable controlled evaluation as context length scales, it spans four context budgets from 8K to 64K tokens. 

Across 15 LALMs, none exceeds 40\% overall accuracy at 32K: models retain \emph{what} was said far better than \emph{who} said it, \emph{how}, or \emph{what} was audible. Accuracy declines as history grows across all evidence types, and error modes differ qualitatively across categories. Our contributions are:
\begin{itemize}[leftmargin=2em, nosep]
\item \textbf{Taxonomy.} A principled taxonomy of spoken conversational memory that jointly characterizes the acoustic evidence to be remembered and the operations.
\item \textbf{Benchmark.} VoxMem, a multi-session benchmark of 3,196 quality-controlled instances, stratified across four context budgets (8K--64K tokens) for controlled study of context length.
\item \textbf{Evaluation.} A systematic evaluation of 15 LALMs revealing structured failures in acoustic memory, with gaps that vary by evidence type, memory operation, and context length.
\item \textbf{Analysis.} Fine-grained error analysis showing that failure modes differ qualitatively across question types, pointing to distinct underlying deficiencies rather than a shared bottleneck.
\end{itemize}
These findings suggest that reliable spoken conversational memory requires more than longer audio contexts, and we provide the taxonomy and benchmark to measure and address these gaps.

\section{Related Work}\label{sec:related}

\textbf{Acoustic evidence and memory operations.}
Spoken memory requires retaining information across two dimensions, acoustic evidence such as 
who said something, how it was said, and what could be heard, and memory operations including retrieval, evidence integration, temporal tracking, etc. 
Existing spoken benchmarks cover both dimensions only partially and ad hoc (Table~\ref{tab:benchmark_comparison_full}): SpokenWOZ and ContextDialog~\citep{si2023spokenwoz,kim2025contextdialog} test only on lexical content of what was said, and others~\citep{he2025audiomarathon,yang2026longspeech,ye2026voicegiraffe,du2025mtalkbench,gosai2026audiomultichallenge,cheng2026voxinfinity} cover only a few selected acoustic evidence, while testing only on retrieval and integration. As a result, they mainly evaluate memory for what was said rather than for the audio, and such heterogeneous evaluations make it hard to tell whether failures arise from a particular evidence type (e.g., whether lexical failures extend to audio~\citep{xiao2026can}), a memory operation, or their interaction. 
VoxMem fills this gap with a principled two-dimensional taxonomy spanning four evidence types and four memory operations, evaluated across all valid combinations.


\textbf{From single-session to multi-session spoken memory.}
Spoken assistants interact with users across temporally separated sessions in which speakers recur and earlier states are updated, so their memory must be evaluated over multi-session histories.
All benchmarks in Table~\ref{tab:benchmark_comparison_full} instead use a long-form single recording or dialogue per instance.
Long-form benchmarks ask about one extended recording~\citep{ahia2025blab,he2025audiomarathon,luo2026chronosaudio,yang2026longspeech,ye2026voicegiraffe},
measuring understanding of the given audio rather than retention across interactions.
Conversational benchmarks test recall within one continuous dialogue~\citep{kim2025contextdialog,si2023spokenwoz,deng2025multibench,du2025mtalkbench,gosai2026audiomultichallenge} of three to eight turns in Audio MultiChallenge. Multi-session histories pose an additional challenge absent from single-session settings: they contain competing sessions on the same topic with different details and many unrelated sessions, requiring a model to locate and bind target evidence rather than rely on topic matching~\citep{jang2023conversation}. VoxMem is, to our knowledge, the first spoken benchmark built on multi-session histories, constructed from evidence, competing, and unrelated sessions across 20 topic families, and evaluated with each question held fixed across context lengths to isolate the effect of history length.
\section{The VoxMem Benchmark}

\begin{figure*}[t!]
    \centering

    \begin{subfigure}[t]{0.417\textwidth}
        \vspace{0pt}
        \centering
        \includegraphics[
    height=0.19\textheight,
    keepaspectratio]{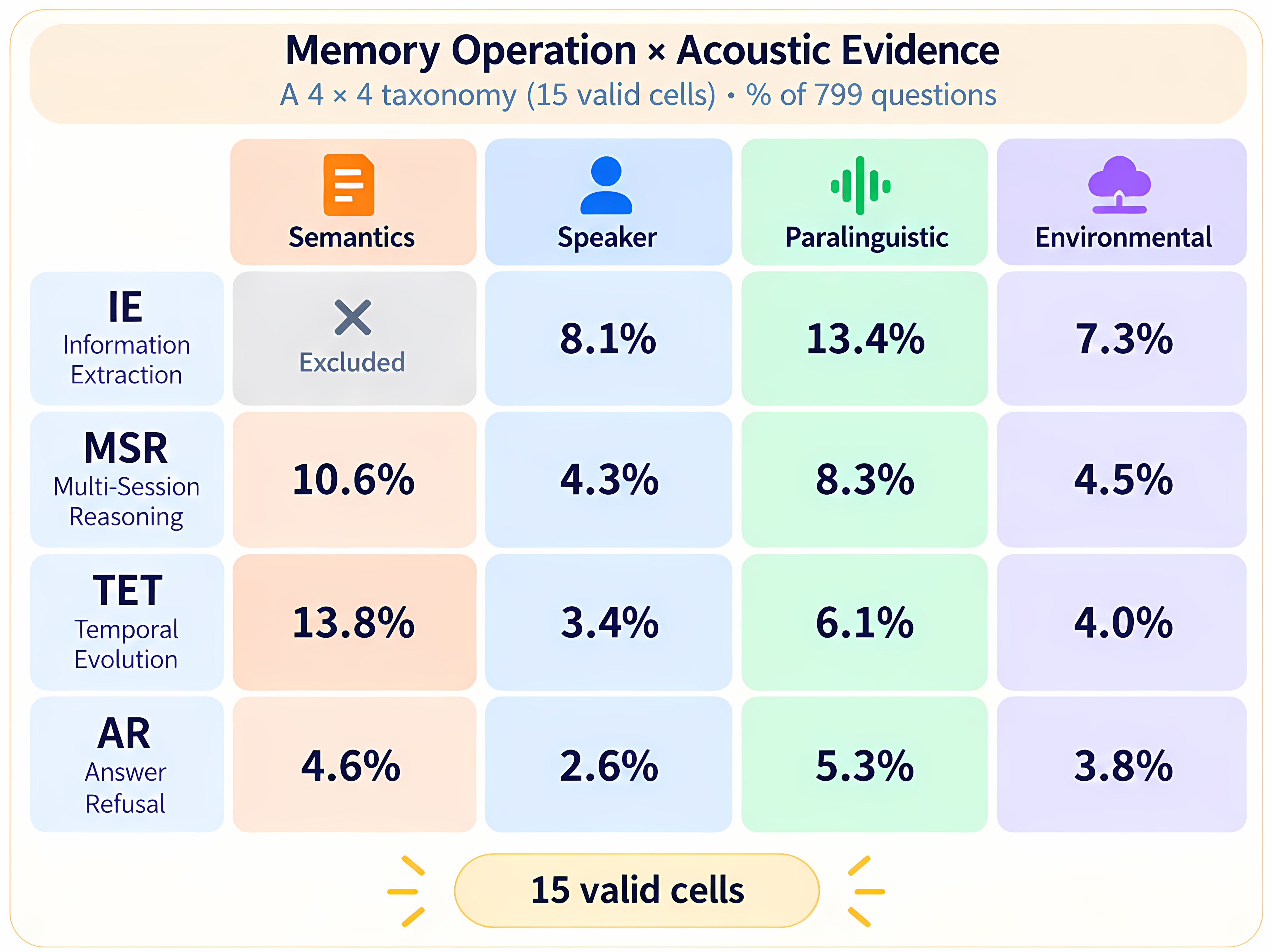}
        \caption{Taxonomy matrix.}
        \label{fig:taxonomy_matrix}
    \end{subfigure}
    \hfill
    \begin{subfigure}[t]{0.268\textwidth}
        \vspace{0pt}
        \centering
        \includegraphics[
    height=0.19\textheight,
    keepaspectratio
]{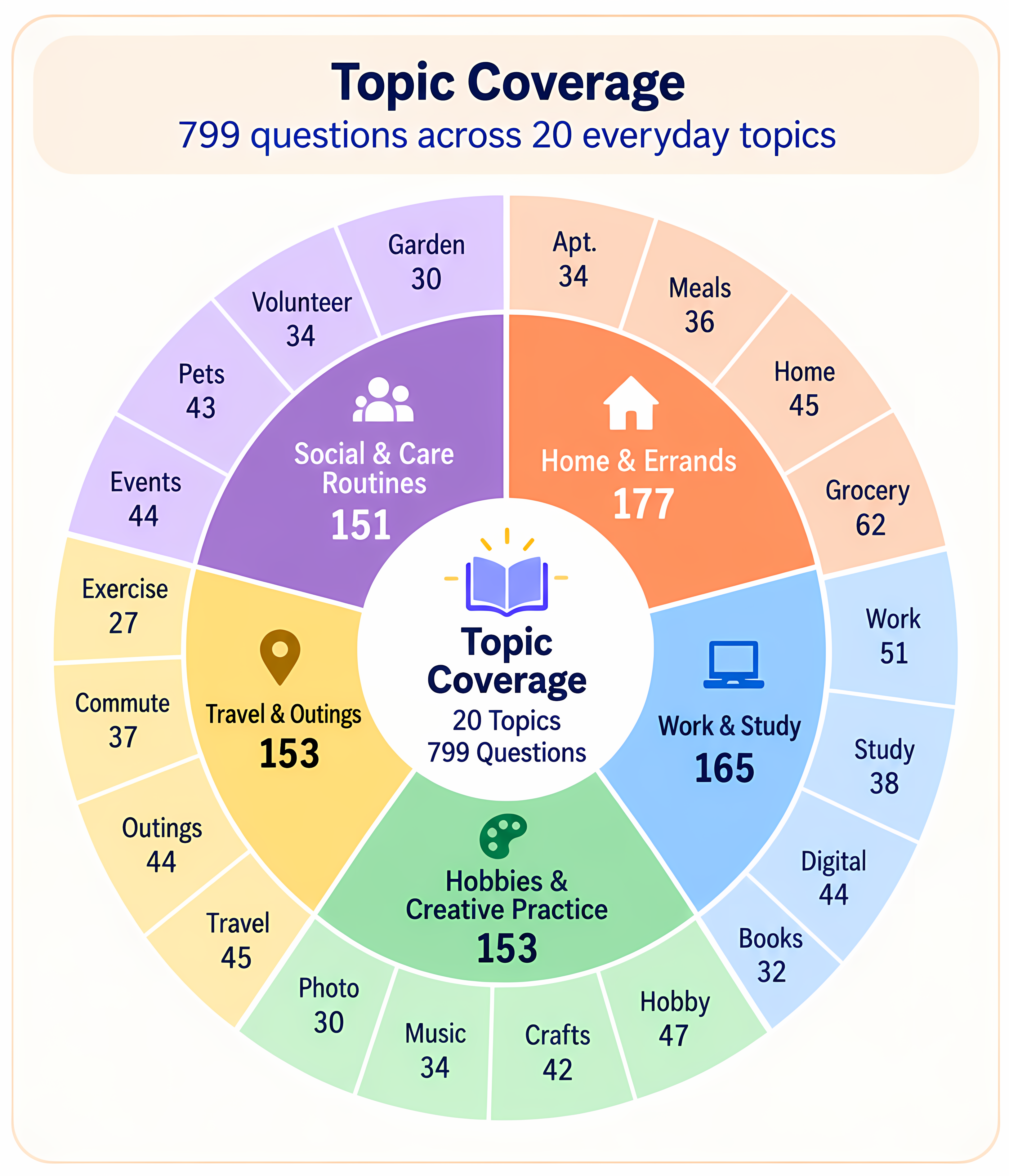}
        \caption{Topic coverage.}
        \label{fig:topic_coverage}
    \end{subfigure}
    \hfill
\begin{subtable}[t]{0.3\textwidth}
    \vspace{0pt}
    \centering

    \resizebox{0.183\textheight}{!}{%
        \begin{tabular}{@{}lr@{}}
            \toprule
            \textbf{Statistic} & \textbf{Value} \\
            \midrule
            Questions                 & 799 \\
            Instances  & 3196 \\
            Sessions & 34{,}743 \\
            Evidence sessions         & 1{,}326 \\
            Avg.\ turns/session       & $\sim$9 \\
            Avg.\ clips/session       & $\sim$4.4 \\
            Tokens/clip               & $\sim$190 \\
            \midrule
            \multicolumn{2}{c}{
                \textit{Controlled scaling: 8K $\rightarrow$ 64K}
            } \\
            Sessions/instance         & 9 $\rightarrow$ 66 \\
            Clips/instance            & 40 $\rightarrow$ 288 \\
            Audio/instance (min)      & 2.5 $\rightarrow$ 20.0 \\
            \bottomrule
        \end{tabular}%
    }

    \caption{Dataset statistics.}
    \label{tab:dataset_stats}
\end{subtable}
\iclrvspace{-2mm}
    \caption{
        \textbf{Overview of the VoxMem benchmark.}
        \textbf{(a)} Memory operations and acoustic evidence form a
        $4\times4$ taxonomy with 15 valid combinations; percentages indicate
        the fraction of the 799 questions in each cell.
        \textbf{(b)} The benchmark covers 20 everyday topics, grouped into
        five broader categories.
        \textbf{(c)} Dataset-scale statistics and controlled context expansion
        from 8K to 64K.
    }
    \label{fig:benchmark_overview}
    \iclrvspace{-5mm}
\end{figure*}



VoxMem establishes a two-dimensional framework for spoken conversational memory evaluation and contributes a benchmark dataset designed to probe both axes systematically. 

\subsection{Two-dimensional Taxonomy} 

As shown in Figure~\ref{fig:benchmark_overview}(a), we define a two-dimensional taxonomy spanning \textit{acoustic evidence} and \textit{memory operations}. \textit{Memory operations} capture how remembered information must be used to answer a later question, and \textit{acoustic evidence} specifies what must be remembered from past speech: what was said (speech semantics), who said it (speaker identity), how it was said (paralinguistic cues), or what could be heard (environmental sound). 


\textbf{Acoustic Evidence.}
The acoustic evidence dimension probes memory across four aspects of spoken interaction, capturing not only what was said but the full richness of the spoken signal.
\textit{Speech semantics} targets memory for \textit{what} was said, such as facts, quantities, plans, preferences, decisions, and explicit updates. As this information is fully recoverable from the words alone, it serves as the transcript-sufficient baseline for semantic memory.
\textit{Speaker identity} targets memory for \textit{who} said what: one assistant serves multiple users of equal status, and answering requires the model to associate previously stated information with the voice of the speaker who produced it — a link absent from any transcript.
\textit{Paralinguistic cues} target memory for \textit{how} something was said, covering vocal states and delivery characteristics such as hesitation, surprise, emphasis, speaking rate, and laughter.
\textit{Environmental sound} targets memory for \textit{what was audible} around the speaker, covering non-speech sound events such as traffic, alarms, machinery, household sounds, and weather.
The latter three types are audio-native by construction: the information necessary to answer is absent from the transcript, making them a direct probe of acoustic memory.

\textbf{Memory Operations.}
VoxMem defines four memory operations, each capturing a distinct way in which evidence distributed across sessions must be used, as shown in Figure~\ref{fig:taxonomy_matrix} with examples in Figure~\ref{fig:example}.
Detailed subtypes are provided in Appendix~\ref{app:operation_subtypes} and Table~\ref{tab:operation_subtypes}. 
\begin{itemize}[leftmargin=2em, nosep]
\item \textbf{Information extraction (IE)} evaluates whether the model can retrieve information from a specific past session.
It is designed so that acoustic evidence is always part of the retrieval, in one of two directions: \textit{context-to-cue retrieval} asks for an acoustic attribute of a session identified by what was said, and \textit{cue-to-fact association} asks for what was said in a session identified by an acoustic cue, such as the voice of the person asking (Figure~\ref{fig:example}(a)).
\item \textbf{Multi-session reasoning (MSR)} evaluates whether the model can combine evidence from several sessions, for example by counting, matching, or comparing it; Figure~\ref{fig:example}(b) asks whether two past conversations share the same background sound.
The answer depends on which sessions are involved but not on their order.
\item \textbf{Temporal evolution tracking (TET)} evaluates whether the model can follow how a piece of information changes across sessions.
We call this information a \textit{state}: a stated plan, which user is taking part, a speaker's tone of voice, or the background sound (Figure~\ref{fig:example}(c)).
A TET question asks for the \textit{latest state}, the \textit{state at a specified earlier point}, or the \textit{full sequence of changes}, so its answer depends on the order of sessions.
For acoustic evidence, a change is not announced in words: the user's tone or the surrounding sound is simply different in a later session, so the model must detect the change from the audio.
\item \textbf{Answer refusal (AR)} evaluates whether the model abstains when the history does not support a unique answer.
AR items are derived from existing IE, MSR, and TET questions by systematically removing or neutralizing the evidence required to answer them (Table~\ref{tab:ar_derivation}).
In speech, the evidence can be neutralized while the words remain, for instance by removing the acoustic cue or the link between a fact and the voice of the person asking, so the model must notice that the acoustic information needed for the answer is gone.
\end{itemize}

\begin{figure*}[t!]
    \centering
    \includegraphics[width=0.8\linewidth]{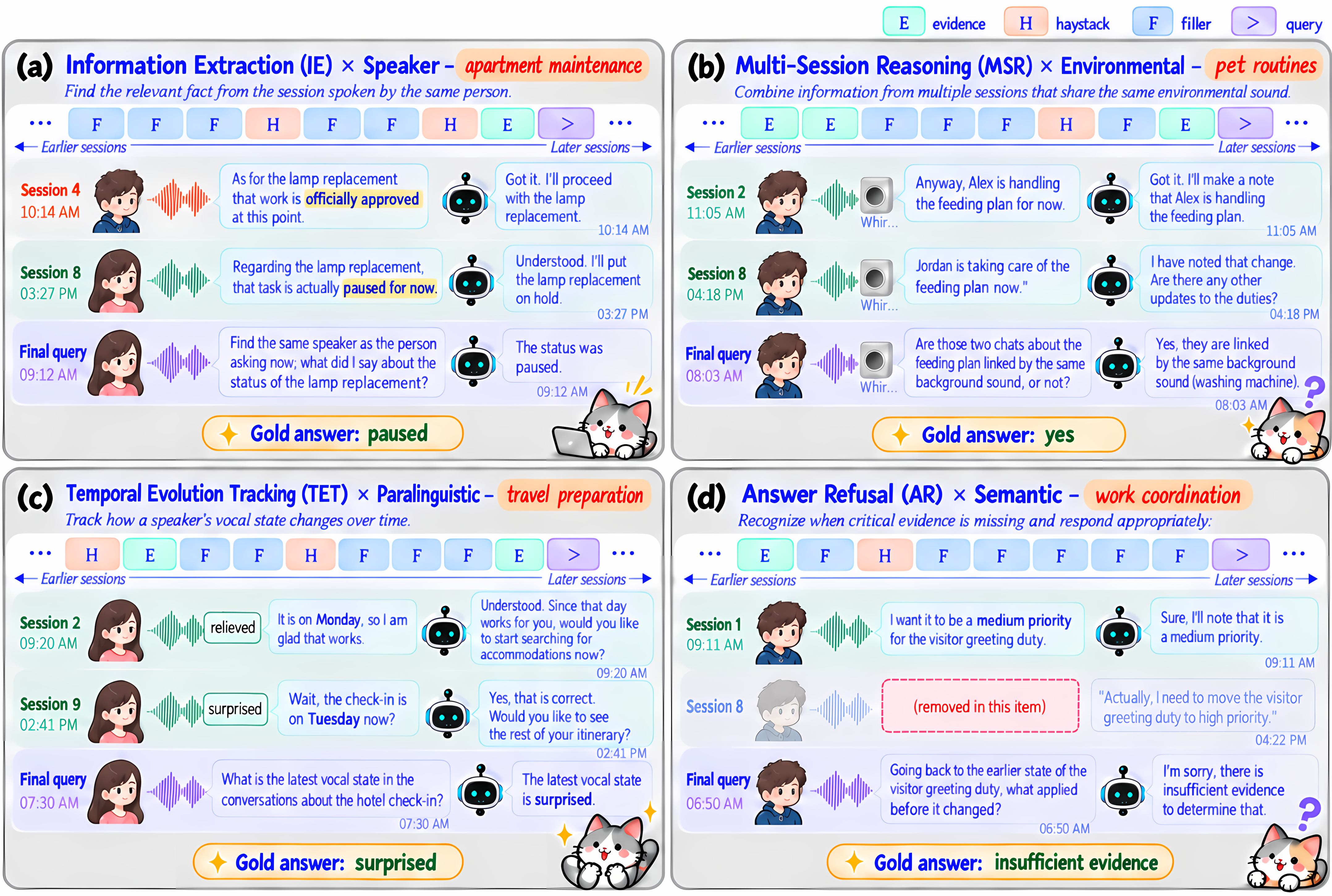}
    \iclrvspace{-2.5mm}
    \caption{\textbf{Representative VoxMem examples.} (a) IE $\times$ Speaker: recall a fact from the querying speaker; (b) MSR $\times$ Environmental: link sessions sharing a background sound; (c) TET $\times$ Paralinguistic: track a changing vocal state; (d) AR $\times$ Semantics: abstain when evidence is missing.}
    \label{fig:example}
    \iclrvspace{-5mm}
\end{figure*}



\subsection{Benchmark Construction}\label{sec:construction}
We construct VoxMem along the two-dimensional taxonomy, covering 15 evaluation scenarios and leaving out information extraction over speech semantics. Each multi-session conversation is composed of three types of sessions designed to ensure both naturalism and controlled scenarios. \textit{Evidence sessions} contain the information required to answer the question.  \textit{Haystack sessions} introduce plausible but misleading content on the same topic, ensuring the model cannot answer by topic matching alone and must reason over the full acoustic and semantic context. \textit{Filler sessions} contain ordinary conversations unrelated to the question, extending the history to the target context length.

Figure~\ref{fig:example}(a) illustrates this structure across memory types. In scenario A, a user asks what they said earlier about a lamp replacement; the answer, ``paused'', appears in one prior session spoken in that user's voice — the \textit{evidence} session. A second session, spoken by a different user, also discusses lamp replacement but with different content, forming the \textit{haystack}: the model cannot simply retrieve the only session on the topic but must identify the correct speaker. The remaining \textit{filler} sessions pad the history with unrelated content; Appendix~\ref{app:assembly} details each session type.

\begin{figure*}[t!]
    \centering
    \iclrarxiv{\includegraphics[width=0.8\linewidth,height=0.2\textheight,keepaspectratio=false]{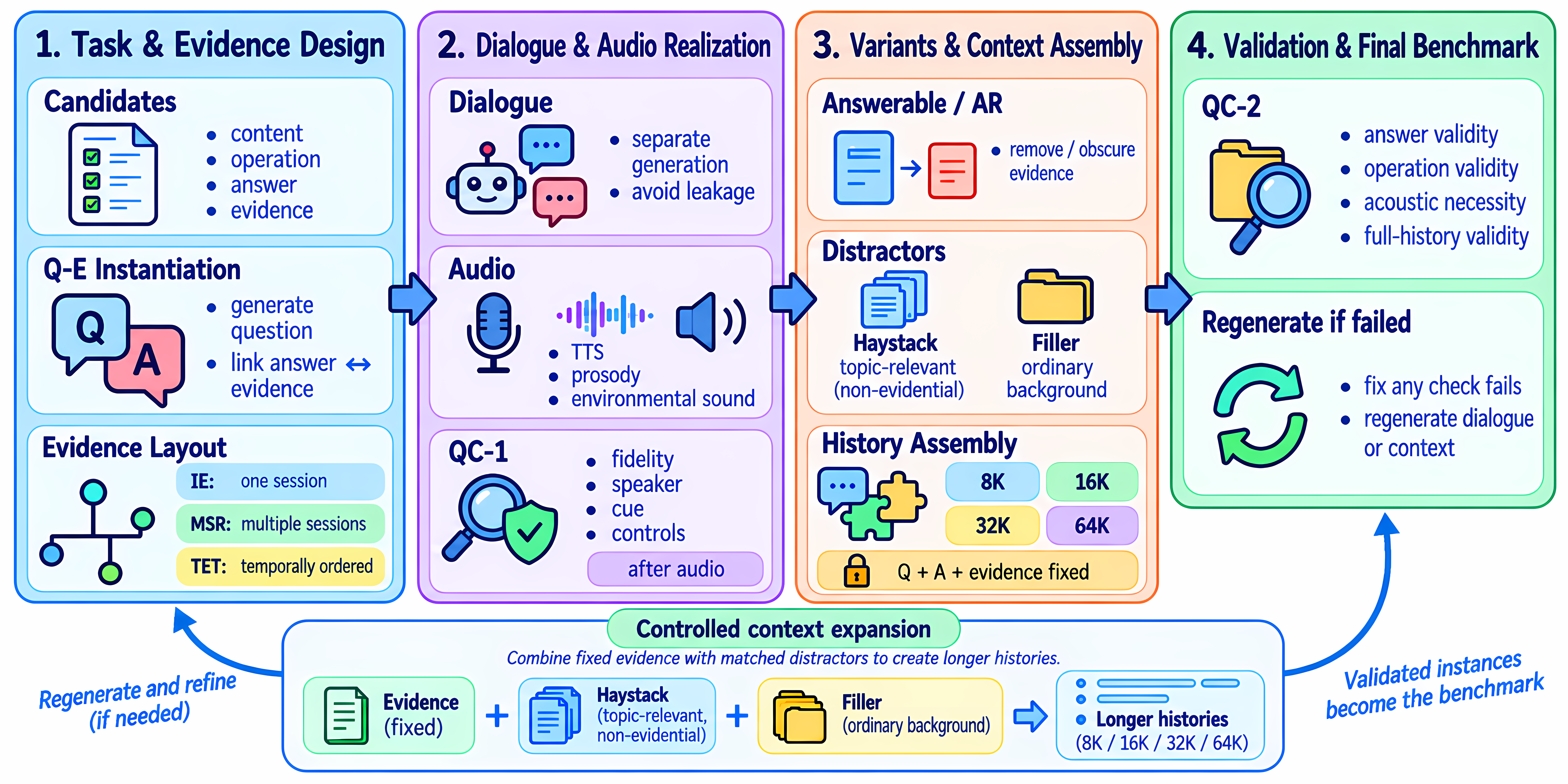}}%
              {\includegraphics[width=0.8\linewidth]{workflow_enhanced.png}}
    \iclrvspace{-2mm}
    \caption{VoxMem construction pipeline: questions are planned from an operation, evidence type, and answer; evidence is realized as spoken dialogue with controlled acoustic cues; histories are built at 8K--64K with question and evidence fixed; multi-stage quality control validates each step.} 
    \label{fig:workflow}
    \iclrvspace{-6mm}
\end{figure*}

Every item follows the same pipeline (Figure~\ref{fig:workflow}).
We first decide the question, its answer, and the evidence it needs, and write the evidence and haystack sessions as text dialogues; filler sessions take their text from InstructS2S~\citep{fang2025llamaomni}, an existing corpus of instruction-following dialogues.
We then synthesize every session with the same TTS system and the voices of the history's users, so that neither voice nor recording quality reveals a session's role, and check each session before assembling them into histories of four lengths.


\textbf{Building evidence sessions.}
Construction follows a three-stage pipeline (Figure~\ref{fig:workflow}). During the first stage of \emph{planning}, each item begins with a structured plan specifying the question, its gold answer, and the required evidence, before any dialogue or audio is written. Evidence is distributed across sessions according to the operation type: a single session for IE, multiple for MSR, and an ordered sequence for TET. The plan also enforces validity constraints: the answer must be unique and, for audio-native items, unrecoverable from the transcript alone. We generate over 30,000 plans, with questions rendered into natural language via Gemini-3.7-Flash and GPT-5.6-Luna.

During the second phase of \emph{dialogue writing}, each evidence session is written as a user--assistant dialogue. To prevent answer leakage, the two sides are authored independently, neither referencing the speaker, vocal tone, or background sounds, so the answer can only be recovered from audio.

The last stage of \textit{speech synthesis} generates the spoken conversation. User turns are synthesized using Higgs-TTS-3~\citep{bosonai_higgs_audio_tts_v3_2026}, with each user assigned a fixed VCTK voice~\citep{vctk} across all sessions. Paralinguistic cues are introduced via style controls and environmental sounds from ESC-50~\citep{piczak2015esc} are mixed in at 10dB SNR. For every such turn, a \textit{paired rendition}, identical in words and voice but with the cue removed, is retained for quality control (Section~\ref{sec:qc}). In the final input, user turns are audio and assistant turns are text, with session boundaries and timestamps.
Because assistant fixed replies should carry no answer-critical acoustic evidence by design, we provide them as text, which removes no information the questions depend on and allows us to evaluate LALMs that accept audio input but do not generate speech; under this setting, we can give every model an identical history.

\textbf{Adding haystack and filler sessions.}
Haystack sessions resemble the evidence but do not yield the answer: some are evidence sessions borrowed from other questions on a compatible topic, and others are written specifically for the question, matching its topic or acoustic context while providing a different value or omitting the answer entirely. To verify their difficulty, a text-only classifier cannot distinguish evidence from haystack sessions (51--56\% accuracy, near the 47--49\% shuffled-label baseline), confirming that the model must use the question, not surface patterns, to locate the evidence. Filler sessions are drawn from InstructS2S~\citep{fang2025llamaomni} and trimmed to match VoxMem session lengths. Every added session is verified not to supply an alternative answer to any item in the benchmark, and no session is reused excessively (Appendix~\ref{app:haystack}).


\textbf{Assembling histories of four lengths.}
To enable controlled evaluation as context grows, each item is embedded in histories at four lengths: 8K, 16K, 32K, and 64K tokens, measured with a Whisper encoder to provide a unified length scale across models (approximately 2.5 to 20 minutes of audio). Longer histories strictly extend shorter ones, preserving all sessions from the shorter history and adding only more haystacks and fillers. This ensures the question, answer, and evidence remain identical across lengths, so any performance drop can be attributed to the growing context rather than a change in the question itself. To prevent positional bias, evidence and haystack sessions are distributed uniformly throughout the history so that neither position nor timestamp signals which session holds the answer (Appendix~\ref{app:assembly}). Sessions whose relative order is question-relevant preserve that order to maintain TET validity. AR items reuse the histories of their source items, modifying only the evidence to ensure no unique answer remains recoverable, keeping the surrounding context identical so that the only change is the availability of evidence.

\subsection{Quality Control}
\label{sec:qc}


Construction errors can cause an item to test something other than its intended cell: a cue may be imperceptible, an answer may not be unique, an audio-native item may be answerable from the transcript, or the assembled history may introduce a spurious answer. These failure modes arise at different construction stages and require validation at two levels. All model-based checks use Gemini-3.7-Flash, which is not among the evaluated models, ensuring no model is assessed on items it helped filter. Items that fail are revised and rechecked; those that still fail are discarded.

\begin{wraptable}{r}{0.45\linewidth}
\vspace{-4mm}
\centering
\small
\caption{
Acoustic-evidence validation with Gemini-3.7-Flash on all 669 answerable questions at 8K. Transcript Only replaces each user turn with its exact synthesis text (all else fixed); $\Delta$: drop from Full Audio.
}
\label{tab:acoustic_validation}
\setlength{\tabcolsep}{3pt}
\begin{tabular}{@{}lrrrr@{}}
\toprule
\makecell[lb]{Evidence\\Type} & $n$ & \makecell[rb]{Full\\Audio} & \makecell[rb]{Transcript\\Only} & $\Delta$ \\
\midrule
Speaker          & 126 & \textbf{69.8} & 10.3 & \textbf{59.5} \\
Paralinguistic   & 222 & \textbf{42.8} &  3.2 & \textbf{39.6} \\
Environmental    & 126 & \textbf{28.6} &  0.8 & \textbf{27.8} \\
Speech Semantics & 195 & \textbf{75.9} & 71.0 & \textbf{4.9} \\
\midrule
All              & 669 & \textbf{54.9} & 23.8 & \textbf{31.0} \\
\bottomrule
\end{tabular}
\vspace{-4mm}
\end{wraptable}
\textbf{Session-level validation (QC-1).} The first level targets failures that can be detected within a single session, before histories are assembled. An audio-language model verifies transcript fidelity, speaker consistency, and the perceptibility of target acoustic cues. Deterministic checks confirm that paralinguistic pairs share the same text and voice reference and environmental pairs keep the clean speech stem.

\textbf{Question-family validation (QC-2).}
The second level targets failures that only emerge once the full history is assembled.
Each answerable question at all four lengths, with its AR variants, undergoes three checks.
\emph{Answer and operation validity} verifies that the evidence uniquely determines the gold answer and that the question requires the intended operation, tested by removing or reordering evidence sessions and confirming the answer changes only as predicted; questions answerable from wording or world knowledge alone are rejected.
\emph{Acoustic necessity} rejects audio-native questions answerable from the transcript or with the cue replaced by its paired rendition.
\emph{Full-history validity} checks each history for leaked answers, alternative answers, and residual AR answerability.

\textbf{Outcome of the acoustic-necessity check.}\label{sec:acoustic-validation}
To confirm that this filter holds on the final benchmark, we compare the full-audio and transcript-only accuracy of Gemini-3.7-Flash on all 669 answerable questions at 8K, where transcript-only input replaces each user turn with its synthesis text.
With speech semantics as a transcript-sufficient reference, transcript-only input reduces accuracy on audio-native questions from 46.2\% to 4.4\%, but on speech semantics only from 75.9\% to 71.0\% (Table~\ref{tab:acoustic_validation}).
Thus, the retained audio-native questions cannot be answered from an oracle transcript.

\section{Evaluation and Analysis}

\subsection{Experimental Setup}

\begin{figure}[t!]
    \centering
    \includegraphics[width=\linewidth]{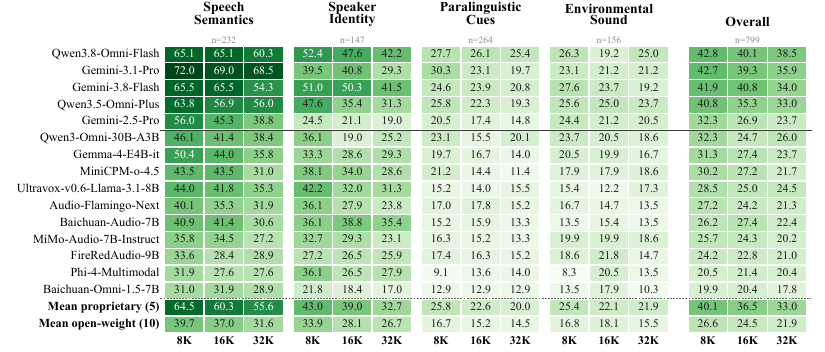}
    \caption{Spoken-memory performance across evidence types and history lengths. Accuracy (\%) of 15 LALMs at the 8K, 16K, and 32K reference history budgets, broken down by the answer-critical evidence type. 
    The horizontal rule separates proprietary and open-weight models, with unweighted group means shown at the bottom.}
    \label{fig:result}
    \iclrvspace{-5mm}
\end{figure}

\textbf{Models and Inputs.}  We evaluate 15 LALMs, including ten open-weight and five proprietary models. The open-weight models span audio-specialist and omni-modal architectures; the proprietary models comprise three Gemini variants and two Qwen variants. All models are evaluated through their native audio interfaces. 
Model versions and other details are in Appendix~\ref{app:models}.


\textbf{Evaluation Protocol.}
We evaluate each model at four history lengths, 8K, 16K, 32K, and 64K tokens, where length is measured in Whisper-encoder tokens so that all models share one scale (Section~\ref{sec:construction}).
Each model receives the interaction history followed by the final query and generates a single response, using the same task instructions for all models (Appendix~\ref{app:prompts}).

\textbf{Scoring and Aggregation.}
All questions are open-ended with short answers, such as a word or phrase, a number, yes or no, or an ordered list of states, and AR items expect the model to state that the history does not support an answer.
Because responses are free text, exact string matching would miss correct answers phrased differently, so we score them with an LLM judge: Gemini-3.7-Flash reads the question, the gold answer, and the response, and marks the response correct or incorrect under a rule for the question's answer type, with refusals attributed to capability or policy counted as incorrect.
Re-judging
responses with GPT-5.6-Luna yields $\kappa = 0.95$ agreement.
We report accuracy, the fraction of questions judged correct, for each evidence type, operation, and length; overall accuracy pools answerable and AR items.
Appendix~\ref{app:judge} lists the answer types, the rule for each, and the judge prompt, all of which are released with the benchmark.

\subsection{Acoustic Evidence}\label{sec:res-acoustic}
Figure~\ref{fig:result} summarizes performance across models, acoustic evidence types, and history lengths. 

\textbf{Finding 1: Current LALMs remain far from reliable spoken conversational memory.} 
At the 32K reference budget, even before the longest histories, no evaluated model exceeds 40\% overall accuracy, with the strongest model reaching 38.5\%.
At this budget, the five proprietary models average 33.0\% and the ten open-weight models 21.9\%.
Although model families differ substantially in absolute performance, low accuracy is not confined to a small subset of models: recovering and appropriately using information from multi-session spoken histories remains challenging.

\textbf{Finding 2: Non-lexical acoustic information is substantially less accessible than speech semantics.}
Performance differs much more sharply across the type of information to be remembered.
On average, speech semantics is the most accessible form of historical information, whereas memory for non-lexical acoustic information is substantially weaker.
At 32K, proprietary models average 55.6\% on Speech Semantics, compared with 32.7\% on Speaker Identity, 20.0\% on Paralinguistic Cues, and 21.9\% on Environmental Sound.
The same qualitative separation appears among open-weight models, whose corresponding means are 31.6\%, 26.7\%, 14.5\%, and 15.5\%.
Together with the transcript-only check in Section~\ref{sec:acoustic-validation}, 
these results show that strong lexical performance masks substantial failures to preserve the broader information conveyed by spoken interaction.

\subsection{Memory Operations}\label{sec:res-operations}
\begin{wrapfigure}{r}{0.535\linewidth}
    \vspace{-4mm}
    \centering
    \includegraphics[width=0.98\linewidth]{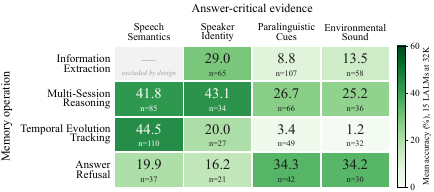}
    \vspace{-2mm}
    \caption{Memory operations interact strongly with the information carried by past speech. Mean accuracy (\%) across 15 LALMs at the 32K reference history budget for each valid combination of memory operation and answer-critical evidence type. Speech-Semantics IE is excluded by benchmark design. Numbers below each score denote the number of evaluation questions in the corresponding cell.}
    \label{fig:op-by-type}
    \vspace{-4mm}
\end{wrapfigure}

Figure~\ref{fig:op-by-type} breaks down performance over the 15 valid operation--evidence combinations at 32K. 

\textbf{Finding 3: Memory difficulty depends on the interaction between the operation and the evidence type.}
Memory operations are not uniformly difficult across evidence types. Multi-Session Reasoning (MSR) is relatively robust: models achieve 41.8\% on Speech Semantics and 43.1\% on Speaker Identity, while still reaching 26.7\% and 25.2\% on Paralinguistic and Environmental evidence—well above the corresponding Information Extraction scores (8.8\% and 13.5\%). Thus, combining evidence across sessions is not necessarily harder than retrieving it from one session. Temporal Evolution Tracking (TET) follows a different pattern. Models track speech semantics well (44.5\%), but performance falls to 20.0\% for Speaker Identity and nearly collapses for Paralinguistic (3.4\%) and Environmental (1.2\%) evidence. The main limitation is not temporal reasoning alone, but maintaining and updating states carried by vocal delivery or ambient sound, possibly reflecting limited stateful modeling in current LALMs.


\textbf{Finding 4: Answer refusal follows a different profile from answering.}
Models abstain more successfully for Paralinguistic and Environmental questions (34.3\% and 34.2\%) than for Speech Semantics and Speaker Identity (19.9\% and 16.2\%), the opposite of the pattern observed in answerable questions. This pattern suggests that models often abstain not because they correctly recognize that the required evidence is absent, but because they struggle to identify or use this acoustic evidence even when it is present. High refusal accuracy on acoustic questions thus reflects a general inability to process acoustic information, not genuine awareness of insufficient evidence.


\begin{figure}[t]
    \centering
    \includegraphics[width=\linewidth]{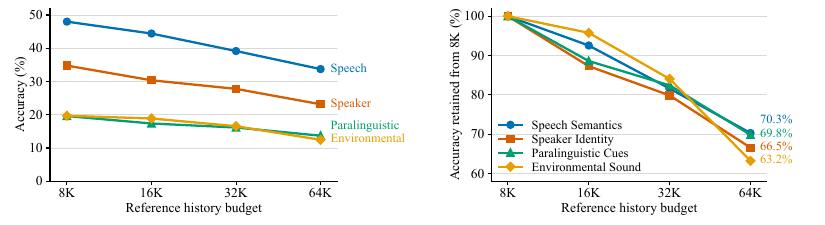}
    \caption{Accuracy scaling by evidence type across context budgets (8K--64K). Left: raw accuracy, showing absolute performance for each evidence type. Right: retention relative to each type's own 8K baseline, isolating the rate of decay from the level difference. }
    \label{fig:scaling}
    \iclrvspace{-4mm}
\end{figure}

\begin{figure*}[t]
    \centering
    \includegraphics[width=\linewidth]{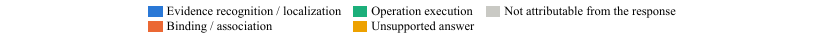}\\[1mm]
    \begin{subfigure}[t]{0.485\textwidth}
        \centering
        \includegraphics[width=\linewidth]{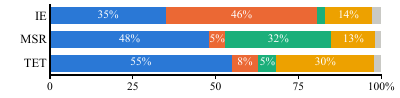}
        \caption{operation}
        \label{fig:a}
    \end{subfigure}\hfill
    \begin{subfigure}[t]{0.485\textwidth}
        \centering
        \includegraphics[width=\linewidth]{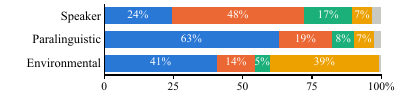}
        \caption{evidence type}
        \label{fig:b}
    \end{subfigure}
    \caption{Error profiles at 64K, attributed with an LLM: composition of incorrect attempted responses (refusals excluded) by (a) answerable memory operation and (b) acoustic evidence type. Full criteria and question breakdowns are in Appendix~\ref{app:err_subtype}.}
    \label{fig:errors}
    \iclrvspace{-5mm}
\end{figure*}


\subsection{Controlled History Length and Error Analysis}\label{sec:res-length}
We study history length and categorize failures via Gemini-3.7-Flash attribution (Figure~\ref{fig:errors}).

\textbf{Finding 5: Access to the same evidence declines as history grows, at different rates across evidence types.}
From 8K to 32K, mean accuracy drops from 40.1\% to 33.0\% for proprietary models and from 26.6\% to 21.9\% for open-weight models. This decline extends to 64K across all four evidence types (Figure~\ref{fig:scaling}). Because VoxMem holds each question and its required evidence fixed across context budgets, these drops can only be attributed to the growing surrounding history: larger context windows do not guarantee stable access to information already present in it.

The rate of decline differs by evidence type. At 64K, models retain 70.3\% of their 8K accuracy on Speech Semantics and 69.8\% on Paralinguistic Cues, compared with 66.5\% for Speaker Identity and 63.2\% for Environmental Sound. Speaker and environmental memory thus degrades fastest, while Paralinguistic memory, despite the lowest absolute accuracy, declines at a rate comparable to Speech Semantics. Low baseline performance and sensitivity to history length are therefore distinct failure modes: models struggle with paralinguistic information at any context length, whereas speaker and environmental memory erodes specifically as history grows.

\textbf{Finding 6: Error profiles differ qualitatively across both evidence types and memory operations.}
Across memory operations, IE errors are dominated by binding and association failures (46\%), meaning models recover an acoustic cue but fail to link it to the correct context. MSR errors split between evidence localization (48\%) and operation execution (32\%), reflecting the additional demands of cross-session comparison and aggregation. TET errors are overwhelmingly localization failures (55\%), with operation-execution errors rare (5\%), indicating that tracking breaks down before reasoning begins. Across evidence types, speaker errors are most often binding failures (48\%), consistent with failures to maintain voice-to-content associations across sessions; paralinguistic errors are dominated by localization failures (63\%), suggesting that answer-critical vocal cues are often not retrieved at all; and environmental errors split between localization (41\%) and unsupported answers (39\%). These profiles show that the three audio-native evidence types are not a single homogeneous difficulty: speaker memory is particularly sensitive to cross-session association, while paralinguistic and environmental memory more often fail before any reasoning takes place.

\section{Conclusion}

We introduced VoxMem, a benchmark for spoken conversational memory organized along two axes of acoustic evidence and memory operations, spanning 15 evaluation scenarios at four controlled context lengths.
Across 15 LALMs, models remember \emph{what was said} far better than \emph{who said it}, \emph{how it was said}, or \emph{what was heard in the environment}. They also track evolving states reliably only when changes are stated in words.
As interaction histories grow, access to previously available evidence consistently degrades, although weaker memory types are not necessarily more sensitive to context length. Importantly, memory failures also differ in nature: speaker-identity errors most often arise from attributing remembered information to the wrong person, whereas paralinguistic errors more often reflect a failure to retain the acoustic cue itself. Thus, simply supporting longer audio contexts is insufficient for persistent spoken interaction. Future systems must preserve non-lexical acoustic information and maintain its associations with speakers, events, and sessions over time.

\bibliography{references_corrected}
\bibliographystyle{iclr2027_conference}

\clearpage
\appendix

\section{Task Definitions and Taxonomy}


This appendix provides the detailed task taxonomy underlying VoxMem. The benchmark separates \emph{what} information must be recovered from past spoken interactions from \emph{how} that information must be used. The former is represented by the evidence-type axis, while the latter is represented by the memory-operation axis. We first summarize the operation subtypes and the scope of each evidence type, and then specify the valid task forms for Information Extraction, Multi-Session Reasoning, and Temporal Evolution Tracking.

\subsection{Memory Operations and Subtypes}
\label{app:operation_subtypes}

The four memory operations characterize different ways in which
historical evidence is used to answer the final query. Information Extraction (IE) identifies and uses information from a
single answer-bearing historical session; Multi-Session Reasoning (MSR) combines evidence distributed across multiple sessions without depending on their temporal order; and Temporal Evolution Tracking (TET) reasons over temporally ordered states whose relative order is answer-critical. Answer Refusal (AR) differs from these three answerable operations: rather than defining an independent family of answerable task subtypes, AR is constructed as a paired counterpart of an IE, MSR, or TET item after the minimum evidence required for a unique answer is removed, neutralized, or made ambiguous.

\begin{table*}[!tbp]
\centering
\small
\renewcommand{\tabularxcolumn}[1]{m{#1}}
\caption{Complete first-level subtype taxonomy. The benchmark contains 30 answerable evidence-type--subtype routes: 6 IE routes, 12 MSR routes, and 12 TET routes. Answer Refusal is generated by paired evidence removal or neutralization and therefore does not introduce an independent subtype family.}
\label{tab:operation_subtypes}
\setlength{\tabcolsep}{6pt}
\begin{tabularx}{\textwidth}{@{} 
  >{\centering\arraybackslash}m{2.0cm} 
  >{\raggedright\arraybackslash}m{3.6cm} 
  >{\raggedright\arraybackslash}X 
  >{\centering\arraybackslash}m{3.6cm} 
@{}}
\toprule
\textbf{Operation} & \textbf{Subtype} & \textbf{One-Sentence Definition} & \textbf{Applicable Evidence Types} \\
\midrule
\raisebox{-2.6\baselineskip}[0pt][0pt]{IE} 
  & Context-to-Cue Retrieval 
  & Uses a semantic or temporal anchor to locate one historical session and then extracts the target speaker, paralinguistic, or environmental attribute from its audio. 
  & Speaker, Paralinguistic, Environmental \\ \addlinespace
  & Cue-to-Fact Association 
  & Uses an acoustic cue to identify one historical session and then retrieves the semantic fact bound to that session. 
  & Speaker, Paralinguistic, Environmental \\
\midrule
  & Counting 
  & Counts occurrences, distinct items, distinct speakers, or sessions satisfying a criterion across multiple evidence sessions. 
  & All four evidence types \\ \addlinespace
MSR 
  & Matching / Resolution 
  & Determines whether cross-session facts, speakers, vocal states, or acoustic events are identical, corresponding, or members of the same group. 
  & All four evidence types \\ \addlinespace
  & Aggregation / Comparison 
  & Combines multiple evidence items through arithmetic, set aggregation, speaker-conditioned aggregation, relative comparison, or dominant-state selection. 
  & All four evidence types \\
\midrule
  & Latest State 
  & Retrieves the most recent valid semantic state, speaker binding, vocal state, or observed acoustic scene. 
  & All four evidence types \\ \addlinespace
TET 
  & Historical State 
  & Recovers the target state at an explicitly specified timestamp, session, or historical stage. 
  & All four evidence types \\ \addlinespace
  & Evolution / Transition 
  & Tracks the ordered sequence, transition pattern, recurrence, or change count of the target evidence across sessions. 
  & All four evidence types \\
\midrule
AR 
  & Paired Derivation 
  & Converts an answerable IE, MSR, or TET item into an unanswerable counterpart by removing, neutralizing, or making ambiguous the minimum necessary evidence. 
  & All four evidence types \\
\bottomrule
\end{tabularx}
\end{table*}

Table~\ref{tab:operation_subtypes} gives the complete first-level subtype taxonomy. Across the three answerable operations, the taxonomy contains 30 evidence-type--subtype routes: six for IE, twelve for MSR, and twelve for TET. Detailed AR transformations are described in Appendix~\ref{app:ar_construction}.

\subsection{Evidence Types and Scope}
\label{app:information_types}

Each VoxMem question is assigned one primary evidence type according to the evidence that is necessary for determining its answer. The four types are Speech Semantics, Speaker Identity, Paralinguistic Cues, and Environmental Sound. This assignment concerns \emph{answer-critical} evidence rather than all information that happens to occur in the audio: non-target acoustic attributes may occur naturally in a session, but they must not systematically predict the gold answer.

Speech Semantics serves as the transcript-sufficient reference condition, covering information explicitly expressed in spoken language. In contrast, answerable Speaker, Paralinguistic, and Environmental questions are constructed so that the required information is not recoverable from lexical content alone. Table~\ref{tab:information_scope} specifies the scope of each evidence type, representative target information, and the corresponding shortcut controls used to keep the design aligned with this distinction.

\begin{table*}[!tbp]
\centering
\small
\renewcommand{\tabularxcolumn}[1]{m{#1}}
\caption{Definitions and scope of the four evidence types. Each question has
one answer-critical evidence dimension; non-target acoustic attributes may
naturally occur but must not systematically predict the answer.}
\label{tab:information_scope}
\setlength{\tabcolsep}{5pt}
\begin{tabularx}{\textwidth}{@{} 
  >{\raggedright\arraybackslash}m{2.5cm} 
  >{\raggedright\arraybackslash\hsize=0.85\hsize}X 
  >{\raggedright\arraybackslash}m{3.6cm} 
  >{\raggedright\arraybackslash\hsize=1.15\hsize}X 
@{}}
\toprule
\textbf{Evidence Type} &
\textbf{Definition} &
\textbf{Representative Information} &
\textbf{Scope and Shortcut Controls} \\
\midrule
Speech Semantics &
Evaluates whether a model can remember, integrate, and update facts, plans,
preferences, and experiences explicitly expressed in spoken language across
multiple sessions. &
Facts, quantities, plans, preferences, decisions, experiences, and explicit
corrections or updates. &
Single-session semantic recall is excluded. Speaker identity, vocal style, and
background sound must not provide the answer. Perfect-transcript input may
remain sufficient by design, making this module a semantic-memory baseline. \\
\addlinespace
Speaker Identity &
Evaluates whether a model can recognize and track voice identity across
sessions and correctly associate people with their statements, actions, and
conversational roles. &
Query-speaker identity, same-speaker matching, speaker--fact binding, and
speaker-conditioned retrieval or aggregation. &
Speaker names or explicit textual identity labels must not reveal the answer.
For speaker tasks, one assistant serves multiple equal-status users, and the
final spoken query is issued by one of the historical users. \\
\addlinespace
Paralinguistic Cues &
Evaluates whether a model can remember and reason over non-lexical vocal cues
that are not fully represented by ordinary transcripts. &
Hesitation, surprise, confidence, tension, excitement, speaking rate, emphasis,
laughter, and other validated vocal states. &
The transcript and assistant reply must not explicitly describe the target
vocal state. Only perceptually reliable and single-turn-recognizable cues are
admitted; fine-grained psychological diagnosis is excluded. \\
\addlinespace
Environmental Sound &
Evaluates whether a model can remember and reason over acoustic scenes and
non-speech events occurring around the conversation. &
Traffic, machinery, alarms, household sounds, weather, and
coarse acoustic scenes. &
Text must not directly name the target sound or location. Questions refer to
the \emph{observed acoustic scene or event}, avoiding unsupported inference
about the user's exact physical location. \\
\bottomrule
\end{tabularx}
\end{table*}

\subsection{Information Extraction}
\label{app:ie_tasks}

Information Extraction evaluates whether a model can recover an
association involving one answer-bearing historical session embedded
within the longer multi-session history. Importantly, IE is not intended
to measure isolated recognition of an acoustic attribute. Each item
requires a two-stage alignment among a retrieval anchor, the relevant
historical session, and the target information.

In \emph{Context-to-Cue Retrieval}, a semantic or temporal anchor first
identifies the relevant historical session, after which the model must
recover an acoustic property of that session, such as speaker identity,
a validated vocal state, or an environmental sound. In \emph{Cue-to-Fact Association}, the direction is reversed: an acoustic cue identifies the relevant historical session, and the model must then recover the semantic fact associated with that session. Speech-Semantics IE is excluded by design. When both the retrieval
signal and the requested information are lexical, a single-session
instance can largely reduce to local speech transcription or semantic
fact retrieval, rather than testing the cross-modal associations that
motivate the IE module. The valid IE task forms and their corresponding
restrictions are summarized in Table~\ref{tab:ie_matrix}.

\begin{table*}[!tbp]
\centering
\small 
\renewcommand{\tabularxcolumn}[1]{m{#1}}
\caption{Information Extraction task matrix. Each IE item contains one answer-bearing historical session and requires a two-stage alignment between a retrieval anchor, the relevant historical session, and the target information. Speech Semantics has no IE category.}
\label{tab:ie_matrix}
\begin{tabularx}{\textwidth}{
  >{\raggedright\arraybackslash\hsize=0.65\hsize}X
  >{\raggedright\arraybackslash\hsize=1.25\hsize}X
  >{\raggedright\arraybackslash\hsize=1.25\hsize}X
  >{\raggedright\arraybackslash\hsize=0.85\hsize}X}
\toprule
\textbf{Evidence Type} &
\textbf{Context-to-Cue Retrieval} &
\textbf{Cue-to-Fact Association} &
\textbf{Key Restriction} \\
\midrule
Speaker Identity &
Use a semantic or timestamp anchor to locate one historical session, then
judge whether its speaker matches the current query speaker or another
reference session. &
Use the spoken query's voice to locate the matching user's historical session,
then retrieve the fact bound to that speaker. &
Prefer relative identity matching over open-ended speaker naming; textual
names and explicit speaker labels are prohibited. \\
\addlinespace
Paralinguistic Cues &
Use a semantic or timestamp anchor to locate one historical session, then
identify its validated vocal state, speaking rate, or emphasis pattern. &
Use a validated paralinguistic cue to locate one historical session, then
retrieve the semantic fact expressed in that session. &
Prefer discrete, perceptually validated cues. Cue, topic, and fact must be
cross-paired so the answer cannot be guessed from content alone. \\
\addlinespace
Environmental Sound &
Use a semantic or timestamp anchor to locate one historical session, then
identify its background event or coarse acoustic scene. &
Use an environmental sound or acoustic scene as the retrieval key, then
recover the semantic fact bound to that session. &
Environment and topic must be cross-paired; text and assistant responses must
not directly mention the target sound or inferred location. \\
\bottomrule
\end{tabularx}
\end{table*}

\subsection{Multi-Session Reasoning}
\label{app:msr_tasks}

Multi-Session Reasoning requires evidence from multiple historical
sessions, but the temporal order of those evidence sessions is not
answer-critical. Operationally, MSR is defined over an unordered set of
answer-critical evidence: permuting the evidence sessions while
preserving their contents should leave the gold answer unchanged.
This property distinguishes MSR from Temporal Evolution Tracking.

The MSR taxonomy contains three first-level subtypes.
\emph{Counting} determines the number of events, items, speakers, or
sessions satisfying a criterion. \emph{Matching / Resolution} determines
whether evidence distributed across sessions refers to the same entity,
speaker, vocal state, event, or group. \emph{Aggregation / Comparison}
combines or compares multiple evidence items to derive a joint answer.
The evidence type determines what serves as the retrieval and
grouping signal, and therefore which historical evidence contributes
to the operation. Table~\ref{tab:msr_matrix} specifies the corresponding
task forms for all four evidence types.

\begin{table*}[!tbp]
\centering
\small 
\renewcommand{\tabularxcolumn}[1]{m{#1}}
\caption{Multi-Session Reasoning matrix. Evidence-session order is not
answer-critical: permuting the evidence sessions should not change the gold
answer. Every item requires evidence from multiple sessions.}
\label{tab:msr_matrix}
\begin{tabularx}{\textwidth}{
  >{\raggedright\arraybackslash\hsize=0.60\hsize}X
  >{\raggedright\arraybackslash\hsize=1.05\hsize}X
  >{\raggedright\arraybackslash\hsize=1.05\hsize}X
  >{\raggedright\arraybackslash\hsize=1.30\hsize}X}
\toprule
\textbf{Evidence Type} &
\textbf{Counting} &
\textbf{Matching / Resolution} &
\textbf{Aggregation / Comparison} \\
\midrule
Speech Semantics &
Count event occurrences, distinct objects, or facts satisfying a semantic
criterion across sessions. &
Resolve whether differently worded references denote the same object, event,
plan, or record. &
Sum quantities, merge sets, compare attributes, or derive a joint answer from
distributed spoken facts. \\
\addlinespace
Speaker Identity &
Count only the sessions belonging to the query speaker, or count distinct
speakers satisfying a shared semantic criterion. &
Perform same-speaker matching, query-to-history identity resolution, or
speaker clustering across sessions. &
Aggregate or compare only the facts bound to a target speaker. Speaker identity
must change the selected evidence set and therefore the answer. \\
\addlinespace
Paralinguistic Cues &
Count sessions or utterances that exhibit a validated vocal state, such as
hesitation or excitement. &
Match, contrast, or group sessions by vocal state or emphasis pattern. &
Compare groups, identify a dominant or majority vocal state, or select the
more confident/hesitant set. Continuous ``emotion arithmetic'' is excluded. \\
\addlinespace
Environmental Sound &
Count occurrences or distinct categories of sound events or acoustic scenes. &
Match or group sessions containing the same sound event, event category, or
coarse acoustic scene. &
Aggregate observed sound categories, compare their frequencies, or identify
the dominant acoustic scene. Uncalibrated loudness arithmetic is excluded. \\
\bottomrule
\end{tabularx}
\end{table*}

\subsection{Temporal Evolution Tracking}
\label{app:tet_tasks}

In contrast to MSR, Temporal Evolution Tracking is defined over an
ordered sequence of historical states. The temporal relationships among
the answer-critical evidence are themselves part of the task:
changing their ordering or temporal anchors may change the gold answer.
Session timestamps provide the default temporal anchors, while
Speech-Semantics items may additionally refer to explicit temporal
expressions contained in the historical speech.

TET contains three first-level subtypes.
\emph{Latest State} asks for the most recent valid state after one or
more historical updates. \emph{Historical State} asks for the state at
a specified earlier timestamp, session, or stage rather than defaulting
to the current state. \emph{Evolution / Transition} asks about the
ordered trajectory itself, including changes, recurrence, appearance,
disappearance, or transition patterns across sessions.

The interpretation of ``state'' depends on the answer-critical
evidence type. For Speech Semantics it may be an explicitly stated
fact, preference, plan, or decision; for Speaker Identity it concerns
the speaker associated with a session or the temporal pattern of speaker
participation and binding; for Paralinguistic Cues it is a
validated vocal state or vocal change; and for Environmental Sound it
is an observed acoustic scene or event. In particular, Speaker TET does
not assume that a person's identity itself changes over time; it tracks
which speaker is associated with the relevant interaction or how speaker
participation and binding evolve across sessions.

Table~\ref{tab:tet_matrix} summarizes the valid TET task forms across
the four evidence types.

\begin{table*}[!tbp]
\centering
\small 
\renewcommand{\tabularxcolumn}[1]{m{#1}}
\caption{Temporal Evolution Tracking matrix. Unlike MSR, TET is order-sensitive:
permuting evidence timestamps may change the answer. Session timestamps provide
the default temporal anchors; Semantic TET may additionally use explicit time
expressions spoken in the historical audio.}
\label{tab:tet_matrix}
\begin{tabularx}{\textwidth}{
  >{\raggedright\arraybackslash\hsize=0.60\hsize}X
  >{\raggedright\arraybackslash\hsize=1.05\hsize}X
  >{\raggedright\arraybackslash\hsize=1.05\hsize}X
  >{\raggedright\arraybackslash\hsize=1.30\hsize}X}
\toprule
\textbf{Evidence Type} &
\textbf{Latest State} &
\textbf{Historical State} &
\textbf{Evolution / Transition} \\
\midrule
Speech Semantics &
Recover the latest valid fact, preference, plan, decision, or state after
successive spoken updates. &
Recover the semantic state at a specified date, session, or update stage,
rather than defaulting to the current state. &
Track the full update chain, change direction, recurrence, or number of
semantic state changes. \\
\addlinespace
Speaker Identity &
Identify the speaker bound to the most recent relevant session or determine
whether that speaker is the current query user. &
Identify the speaker at a specified timestamp and compare that identity with
the query speaker or another reference session. &
Track speaker appearance, participation, or handoff order. This does
\emph{not} model a person's identity as changing over time. \\
\addlinespace
Paralinguistic Cues &
Recover the main user's latest relevant vocal state or emphasis pattern. &
Recover the main user's vocal state at a specified timestamp, session, or
stage. &
Track an ordered sequence of validated vocal states, speaking-rate changes,
emphasis shifts, or state recurrence. \\
\addlinespace
Environmental Sound &
Recover the latest observed acoustic scene or sound event in the relevant
history. &
Recover the observed acoustic scene or event at a specified timestamp or
session. &
Track the ordered transition, disappearance, recurrence, or first/last
appearance of acoustic scenes or sound events. \\
\bottomrule
\end{tabularx}
\end{table*}

\section{Benchmark Construction and Validation Details}


This appendix provides the construction and validation details omitted
from the main text. We first describe how structured question
specifications are instantiated as spoken dialogues, then detail the
acoustic controls, answer-refusal derivation, distractor and session-reuse
rules, and controlled history assembly. Quality control is described in
\S\ref{sec:qc}.

\subsection{Question and Evidence Specification}
\label{app:qgen}

Each VoxMem question begins as a structured candidate defined before any
dialogue or audio is generated. A candidate specifies (i) the memory
operation, subtype, and evidence type; (ii) the target information
and the condition selecting the historical session or sessions relevant
to the query; (iii) the operation applied to that selection; (iv) the
answer type and gold value; (v) the supporting facts or acoustic cues
and their session assignments; and (vi) validity requirements such as
answer uniqueness and acoustic necessity.

For example, a Context-to-Cue IE item may specify a semantic anchor
identifying exactly one historical session and ask for the speaker
associated with that session. The structured representation records the anchor, the selected session cardinality, the target speaker slot, the supporting evidence, and the requirement that removing the speaker cue renders the question unanswerable.

\paragraph{Enumeration.}

We issue 12,000 generation requests spanning the 30 valid evidence-type--subtype routes described in Appendix A. Each request is associated with an independent content blueprint and may produce multiple candidate specifications under its generation quota, yielding more than 30,000 candidates before validation. Blueprints cover the 20 topic families used in VoxMem together with the permitted speaker, paralinguistic, and environmental attributes.

\paragraph{Candidate validation.}
Before dialogue generation, deterministic checks verify schema integrity,
route and cardinality consistency, identifier references, permitted
acoustic labels, and the gold answer recomputed from the structured
evidence. Evidence-dependency and shortcut checks are applied to each question
family after history assembly (\S\ref{sec:qc}). Candidates failing any required check are not
admitted to dialogue generation.

\subsection{Spoken Dialogue Generation}
\label{app:dialogue}

Each evidence session is realized as a complete user--assistant dialogue
using separate user and assistant generation passes. This separation
prevents the generation process from copying answer-bearing information
from the user evidence into assistant text.

\paragraph{User generation.}
The user generator receives only the session-local plan, including the
facts assigned to that session, their turn assignments, and constraints
on the surrounding dialogue. It does not receive the final query, the
gold answer, other sessions, distractors, or the corresponding
answer-refusal variant. Required semantic facts are realized in their
designated turns while preserving their intended meaning, and generated
spans are aligned back to the structured evidence.

\paragraph{Assistant generation.}
The assistant generator is evidence-blind: it receives the visible
conversation prefix but not the benchmark evidence specification,
gold answer, acoustic labels, final query, future user turns, or other
sessions. Assistant replies may not repeat, summarize, or confirm
answer-bearing details from the preceding user turn. They are additionally
prohibited from explicitly describing speaker identity, vocal state,
background sound, or other target acoustic attributes.

\paragraph{Audio-native constraints.}
For Speaker, Paralinguistic, and Environmental questions, the user text
must remain compatible with the target acoustic realization without
lexicalizing it. Thus the target speaker, vocal state, or environmental
sound cannot be inferred directly from the canonical transcript.
Table~\ref{tab:gen_constraints} summarizes the lexical constraints used
for the four evidence types.

\begin{table}[!tbp]
\centering
\small
\renewcommand{\tabularxcolumn}[1]{m{#1}}
\caption{\textbf{Generation constraints and evidence localization by evidence type.} Speech Semantics serves as the transcript-sufficient reference. For all audio-native types, strict lexical constraints prevent transcript-based shortcuts, ensuring that answer-critical evidence resides exclusively in the acoustic channel. Evidence localization defines the temporal scope of the relevant audio evidence.}
\label{tab:gen_constraints}
\begin{tabularx}{\textwidth}{
  >{\raggedright\arraybackslash\hsize=0.60\hsize}X
  >{\raggedright\arraybackslash\hsize=1.40\hsize}X
  >{\raggedright\arraybackslash\hsize=1.00\hsize}X}
\toprule
\textbf{Evidence Type} &
\textbf{Lexical Constraint} &
\textbf{Evidence Localization} \\
\midrule
Speech Semantics &
None (transcript-sufficient reference). The transcript must explicitly contain all answer-bearing facts, entities, and state changes. &
Exact semantic span within the target user turn. \\
\addlinespace[3pt]
Speaker Identity &
Strictly prohibits speaker naming, explicit self-introductions, or demographic descriptors; text must remain voice-neutral and compatible with counterfactual speaker assignments. &
Whole audio turn (speaker voice characteristics across the utterance). \\
\addlinespace[3pt]
Paralinguistic Cues &
Strictly prohibits vocal-state labels, descriptive emotion adverbs, or self-reporting (e.g., ``I feel hesitant''); text is speakable in the target style without lexical leakage. &
Whole audio turn, or a designated temporal segment when localized to a specific phrase. \\
\addlinespace[3pt]
Environmental Sound &
Strictly prohibits mentioning ambient sound events or naming physical environments; dialogue must not semantically entail or hint at the acoustic background. &
Whole audio turn (continuous background acoustic scene or discrete sound event). \\
\bottomrule
\end{tabularx}
\end{table}

\begin{table}[!tbp]
\centering
\small
\renewcommand{\tabularxcolumn}[1]{m{#1}}
\caption{\textbf{Audio realization settings.} Paired controls preserve the canonical transcript and non-target acoustic dimensions while varying the answer-critical cue.}
\label{tab:audio_settings}
\begin{tabularx}{\linewidth}{
  >{\raggedright\arraybackslash\hsize=0.65\hsize}X
  >{\raggedright\arraybackslash\hsize=1.35\hsize}X}
\toprule
\textbf{Component} & \textbf{Setting} \\
\midrule
Speech synthesis &
Higgs-TTS-3 (4B), fixed model revision. \\
\addlinespace[2pt]
Audio format &
24\,kHz, 16-bit, mono WAV. \\
\addlinespace[2pt]
Speaker control &
30 fixed VCTK reference speakers; each persona retains the same reference across sessions. \\
\addlinespace[2pt]
Paralinguistic control &
Controlled vocal-state/prosody tags; cued and neutral renditions share the same canonical transcript and voice reference. \\
\addlinespace[2pt]
Environmental assets &
45 selected ESC-50 types. \\
\addlinespace[2pt]
Environmental mixing &
Additive mixing at 10\,dB target SNR; the matched ablation preserves the clean speech stem. \\
\bottomrule
\end{tabularx}
\end{table}

\subsection{Acoustic Realization and Controls}
\label{app:audio}

All user speech is synthesized with Higgs-TTS-3 (4B) using a fixed
generation stack. Speaker identity is controlled through fixed zero-shot
VCTK voice references: each benchmark persona is assigned one reference
speaker and retains that voice across all sessions.

Paralinguistic cues are imposed through synthesis controls while holding
the canonical transcript and speaker reference fixed. For each controlled
paralinguistic realization, a corresponding neutral rendition preserves
the same text and voice and differs only in the target control.

Environmental evidence is introduced by mixing a selected ESC-50 sound
with the clean speech stem at a fixed target SNR of 10\,dB. The matched
environment-removed rendition preserves the clean speech stem. These
paired constructions allow the target acoustic dimension to be removed
without changing the lexical content or non-target speech signal.

\subsection{Answer Refusal Construction}
\label{app:ar_construction}

Each Answer Refusal (AR) item is derived from an admitted answerable IE,
MSR, or TET item by removing, neutralizing, or obscuring the evidence
required for a unique supported answer. The final query and reference
context budget are preserved, while modifications are restricted to the
answer-critical evidence path.

Table~\ref{tab:ar_derivation} summarizes the five derivation rules used for
AR construction. The paired design allows the supported and
insufficient-evidence variants to differ primarily in whether the
required historical evidence is available. Of 166 constructed AR
variants, 130 passed the full validation procedure and are included in
the benchmark.

\begin{table*}[!tbp]
\centering
\small 
\renewcommand{\tabularxcolumn}[1]{m{#1}}
\caption{Paired derivation rules for Answer Refusal. Each AR item is produced
from an answerable source item while preserving the question wording, context
budget, haystack sessions, and filler sessions. The gold
answer becomes a canonical refusal because the remaining history no longer
supports a unique answer.}
\label{tab:ar_derivation}
\begin{tabularx}{\textwidth}{
  >{\raggedright\arraybackslash\hsize=0.65\hsize}X
  >{\raggedright\arraybackslash\hsize=0.90\hsize}X
  >{\raggedright\arraybackslash\hsize=1.25\hsize}X
  >{\raggedright\arraybackslash\hsize=1.20\hsize}X}
\toprule
\textbf{Derivation Rule} &
\textbf{Typical Source Route} &
\textbf{Transformation} &
\textbf{Why the Remaining Context Is Insufficient} \\
\midrule
Complete Evidence Removal &
IE or single-anchor tasks &
Remove the sole answer-bearing evidence session while preserving matched
haystack and filler sessions. &
No remaining session contains the fact or acoustic cue required by the
question. \\
\addlinespace
Partial Evidence Removal &
MSR or TET &
Remove one or more minimally necessary evidence sessions from a multi-session
aggregation or ordered update chain. &
The remaining subset does not determine the count, aggregate, historical
state, or transition sequence. \\
\addlinespace
Cue Neutralization &
Speaker, Paralinguistic, or Environmental routes &
Preserve lexical content but replace, suppress, or neutralize the
answer-critical voice identity, vocal state, or environmental sound. &
Semantic content remains plausible, but the acoustic discriminator needed to
select or interpret the target session is absent. \\
\addlinespace
Binding Removal &
Speaker routes &
Preserve candidate facts while removing the reliable voice-based link between
the facts and the current query speaker. &
Several users' facts remain possible, so the assistant cannot determine which
fact belongs to the person asking the question. \\
\addlinespace
Ambiguity Construction &
IE, MSR, or TET &
Retain multiple equally plausible candidates after removing the unique
disambiguating evidence. &
The history supports more than one answer, so committing to any single content
answer would be hallucination. \\
\bottomrule
\end{tabularx}
\end{table*}

\subsection{Haystack, Filler, and Session Reuse}
\label{app:haystack}

Long histories contain three session roles: \emph{evidence sessions},
\emph{haystack sessions}, and \emph{filler sessions}, as
summarized in Table~\ref{tab:session_roles}. Evidence sessions contain
the answer-critical information specified by the source question.
Haystack sessions are deliberately selected to resemble relevant history
without supporting the gold answer, whereas filler sessions provide
ordinary conversational context without being matched to the target
question.

\paragraph{Haystacks.}
Each validated evidence session belongs to exactly one source question,
but may additionally serve as a haystack for other compatible questions.
A reused session is eligible as a haystack only when it originates from
a different source question, matches the target evidence type and
required voice organization, satisfies the predefined topic-compatibility
rule, and contains no answer-bearing fact from the target question.
Surviving candidates are further screened for gold-answer overlap,
joint-inference shortcuts, alternative uniquely supported answers, and
incompatible speaker, timestamp, or acoustic configurations. In addition to reused haystacks, VoxMem constructs question-specific haystacks as
\emph{same-key distractors}. These sessions share the retrieval cue or selection criterion of the target question but do not support its gold answer. For example, a haystack may match the queried topic or acoustic attribute while carrying a non-target value, or may satisfy the retrieval cue without containing the queried attribute at all. Same-key distractors therefore create competition around the relevant retrieval key rather than serving as generic background context.

\paragraph{Filler sessions.}
Filler sessions are not generated specifically for VoxMem, but are drawn from InstructS2S~\citep{fang2025llamaomni}, an existing corpus of ordinary instruction-following speech dialogues. Before entering the filler pool, candidates are trimmed to match VoxMem session lengths, reducing the extent to which session length alone can reveal session role. They are then screened against the complete set of benchmark atomic facts and gold answers, and are rejected if they state or directly imply protected answer-bearing content. When assigned to a particular history, filler sessions additionally pass compatibility checks for voice organization and timestamps. This global semantic screening allows an admitted filler session to be reused across question families without introducing a known answer path.

\paragraph{Reuse constraints.} Reuse is capped to prevent a small number of sessions from dominating the benchmark. Within each evidence-type--operation--subtype stratum, a haystack session group may be reused at most twice, while a filler session may appear at most 24 times. Paired acoustic renditions of the same underlying session share a single reuse count. These constraints govern session eligibility and reuse; the number and placement of each session role at different context budgets are determined by the history assembly procedure in Appendix~\ref{app:assembly}.  
\begin{table*}[!tbp]
\centering
\small
\renewcommand{\tabularxcolumn}[1]{m{#1}}
\caption{Definitions and shared-pool design for evidence, haystack, and filler sessions. Evidence and haystack sessions share a module-specific pool: a session that is evidence for its source question may serve as haystack for a different compatible question, subject to reuse limits and answer-contamination
checks.}
\label{tab:session_roles}
\begin{tabularx}{\textwidth}{
  >{\raggedright\arraybackslash\hsize=0.55\hsize}X
  >{\raggedright\arraybackslash\hsize=0.95\hsize}X
  >{\raggedright\arraybackslash\hsize=1.10\hsize}X
  >{\raggedright\arraybackslash\hsize=1.15\hsize}X
  >{\raggedright\arraybackslash\hsize=1.25\hsize}X}
\toprule
\textbf{Session Role} &
\textbf{Answer Relation} &
\textbf{Topical / Acoustic Profile} &
\textbf{Pool and Reuse Policy} &
\textbf{Required Validation} \\
\midrule
Evidence &
Contains one or more facts or acoustic cues that are minimally necessary for
the current gold answer. &
Generated for the target module and route; answer-critical information is
embedded naturally rather than highlighted. &
Stored in a module-specific pool. A session is evidence for its source
question and may be reused only a limited number of times across released
instances. &
Answerability, uniqueness, cue validity, semantic fidelity, naturalness, and
absence of transcript or assistant-response leakage. \\
\addlinespace
Haystack &
Does not support the current answer but is intentionally confusable with the
current evidence. &
Matched to the evidence along task-relevant axes: topic and fact type for
Semantic, voice identity structure for Speaker, vocal-state distribution for
Paralinguistic, and sound-event distribution for Environmental. &
Selected primarily from other compatible questions' evidence sessions in the
same module-specific pool. Reuse is capped, and the same atomic fact or gold
answer is excluded. &
Evidence-vs.-haystack classifiers should remain near chance; grouped splits
prevent duplicate-session leakage. Each candidate passes answer-contamination
and alternative-answer checks. \\
\addlinespace
Filler &
Provides no answer-relevant information and mainly expands the long-term
interaction history. &
Broader and more ordinary conversation distribution. It may contain natural
speech variation, but no deliberately structured target cue or evidence
chain. &
Drawn from a separate filler pool. Filler is interleaved
throughout the history rather than confined to one position. &
Check that filler is not trivially separable through recording quality,
speaker, length, TTS settings, or the complete absence of ordinary acoustic
variation. \\
\bottomrule
\end{tabularx}
\end{table*}

\subsection{Controlled History Assembly}
\label{app:assembly}

After evidence, haystack, and filler sessions have passed the eligibility rules above, they are assembled into four nested histories with nominal reference budgets of 8K, 16K, 32K, and 64K. Across all four instantiations, the question, gold answer, and answer-critical evidence remain fixed; only the surrounding interaction history expands.  

\paragraph{Nested context growth.} The amount of matched distraction increases with both the reference budget and the amount of evidence required by the question. Reused haystack allocation is conditioned on the number of evidence sessions for that item and increases progressively across the four tiers. Question-specific same-key distractors follow a nested schedule of up to 1, 2, 4, and 8 sessions at 8K, 16K, 32K, and 64K, respectively. Filler sessions are then used to occupy the remaining context budget. Because the histories are nested, sessions introduced at a shorter budget are retained when constructing the corresponding longer history.  

\paragraph{Session placement.} Evidence and matched haystack sessions follow the same positional policy so that location alone does not identify which sessions carry the answer-critical information. Additional sessions are interleaved throughout the history rather than appended as a separate block. Assembly preserves all temporal relations required by the source question; in particular, the ordering of answer-critical states in TET items is unchanged when distractor and filler sessions are inserted.  

\paragraph{Resulting history scale.} Figure~\ref{tab:dataset_stats} reports the resulting history scale from 8K to 64K: the number of sessions, audio clips, and minutes of audio per instance. The common reference-length measure is used only to control history scale across models; the model-facing question and evidence are unchanged across budgets. 

\paragraph{Text separability of distractors.} A TF--IDF logistic-regression
probe reads only session text and tries to separate evidence from haystack
sessions, with cross-validation splits grouped by question. It reaches
51--56\% accuracy, against 47--49\% when the labels are shuffled,
so the text of a session does not reveal whether it is evidence or haystack,
and a model has to use the question to find the evidence. Filler sessions contain
nothing relevant to any answer, since every filler session is screened against
all benchmark atomic facts and gold answers.

\section{Evaluation Protocol}
\label{app:protocol}


%

This appendix specifies how every model is run and scored. We list the
evaluated systems and their exact versions (\S\ref{app:models}), the input
serialization shared by all models (\S\ref{app:input}), the reference scale on
which history length is measured (\S\ref{app:length}). We then give the scoring rubric
(\S\ref{app:judge}), its cross-validation against a second judge
(\S\ref{app:xjudge}), and the model set used wherever 64K results are reported
(\S\ref{app:sixtyfour}).

\subsection{Evaluated Models and Exact Versions}
\label{app:models}

Table~\ref{tab:c1} lists every evaluated system. Open-weight models are
identified by their Hugging Face repository; the exact revision loaded is recorded
in the code release, because several repositories were updated during the
evaluation window. Proprietary models are identified by the API model
string sent with each request and, where the provider publishes one, the dated
snapshot that string pointed to on the evaluation date. All fifteen systems
receive user turns as audio through their native audio interface; no system is
given transcripts or a speech-recognition front end. Inference settings for
every system are released with the code.

\begin{table*}[!htbp]
\centering
\footnotesize
\setlength{\tabcolsep}{4pt}
\resizebox{\textwidth}{!}{
\begin{tabular}{lll rr}
\toprule
\textbf{Model} & \textbf{Family} & \textbf{Hugging Face Repo / API Model String} & \textbf{LLM} & \textbf{LLM+Enc.} \\
\midrule
\multicolumn{5}{l}{\textit{Audio-specialist models}} \\
\midrule
Audio-Flamingo-Next~\citep{ghosh2026audioflamingonext}        & Audio Flamingo & \texttt{nvidia/audio-flamingo-next-hf}         & 7.6B & 8.3B \\
MiMo-Audio-7B-Instruct~\citep{mimo}     & MiMo           & \texttt{XiaomiMiMo/MiMo-Audio-7B-Instruct}     & 7.6B & 8.4B\textsuperscript{$\ast$} \\
Baichuan-Audio-7B~\citep{baichuan}          & Baichuan       & \texttt{baichuan-inc/Baichuan-Audio-Instruct}  & 7.6B & 8.6B \\
Ultravox-v0.6-Llama-3.1-8B & Ultravox       & \texttt{fixie-ai/ultravox-v0\_6-llama-3\_1-8b} & 8.0B\textsuperscript{$\dagger$} & 8.7B \\
FireRedAudio-9B~\citep{li2026fireredaudio}            & FireRed        & \texttt{FireRedTeam/FireRedAudio}              & 9.0B & 9.6B\textsuperscript{$\S$} \\
\midrule
\multicolumn{5}{l}{\textit{Omni-modal models}} \\
\midrule
Phi-4-Multimodal~\citep{abouelenin2025phi4mini}           & Phi            & \texttt{microsoft/Phi-4-multimodal-instruct}   & 3.8B & 5.6B \\
Gemma-4-E4B-it~\citep{gemma4}             & Gemma          & \texttt{google/gemma-4-E4B-it}                 & 7.5B\textsuperscript{$\P$} & 8.0B\textsuperscript{$\P$} \\
MiniCPM-o-4.5~\citep{cui2026minicpmo}              & MiniCPM        & \texttt{openbmb/MiniCPM-o-4\_5}                & 8.2B & 9.0B \\
Baichuan-Omni-1.5-7B~\citep{li2025baichuanomni}       & Baichuan       & \texttt{baichuan-inc/Baichuan-Omni-1d5}        & 7.6B & 9.3B \\
Qwen3-Omni-30B-A3B~\citep{xu2025qwen3omni}         & Qwen-Omni      & \texttt{Qwen/Qwen3-Omni-30B-A3B-Instruct}      & 30.5B\textsuperscript{$\ddagger$} & 31.7B\textsuperscript{$\ddagger$} \\
\midrule
\multicolumn{5}{l}{\textit{Proprietary APIs}} \\
\midrule
Qwen3.8-Omni-Flash         & Qwen-Omni      & \texttt{qwen3.8-omni-flash}                    & --- & --- \\
Gemini-3.1-Pro             & Gemini         & \texttt{gemini-3.1-pro-preview}                & --- & --- \\
Gemini-3.8-Flash           & Gemini         & \texttt{gemini-3.8-flash}                      & --- & --- \\
Qwen3.5-Omni-Plus          & Qwen-Omni      & \texttt{qwen3.5-omni-plus} (= \texttt{qwen3.5-omni-plus-2026-03-15}) & --- & --- \\
Gemini-2.5-Pro             & Gemini         & \texttt{gemini-2.5-pro}                        & --- & --- \\
\bottomrule
\end{tabular}}
\caption{\textbf{Evaluated systems.} The 15 LALMs evaluated in VoxMem, grouped
into audio-specialist and omni-modal open-weight models and proprietary APIs;
open-weight models are sorted by \textbf{LLM+Enc.} within each group.
\textbf{LLM} is the text backbone, including embeddings and output head;
\textbf{LLM+Enc.} adds all input-side modules (audio and vision encoders,
projectors and adapters) and excludes speech-output modules. Parameter counts
are read from the safetensors headers of each repository.
$^{\ast}$Includes the encoder of \texttt{XiaomiMiMo/MiMo-Audio-Tokenizer}
(0.65B). $^{\dagger}$The checkpoint holds only the audio encoder and projector;
the LLM count is that of \texttt{meta-llama/Llama-3.1-8B-Instruct}.
$^{\S}$Excludes the RedAE and patch modules used for speech generation
(0.99B). $^{\P}$Includes per-layer embeddings; 4.5B effective parameters.
$^{\ddagger}$Mixture-of-experts, about 3B parameters active per token.}
\label{tab:c1}
\end{table*}

\subsection{Input Formatting}
\label{app:input}

Every model receives one conversation followed by one query and returns a
single response. The serialization is identical across models.

A history is a flat, chronologically ordered sequence of turns. A user turn
carries audio; an assistant turn carries text. The first user turn of each
session is preceded by a text block with that session's timestamp, which is the
only session-boundary marker: no separator tokens, headers or session indices
are inserted, so a model must infer session structure from the timestamps. The
final session holds only the spoken query, so the query is also preceded by
its timestamp, which gives the present time.

\begin{lstlisting}[basicstyle=\fontsize{7.5pt}{9pt}\ttfamily, breaklines=true]
  system     You are continuing an ongoing conversation that took place across
             multiple timestamped sessions. User turns are provided as audio,
             and assistant turns are provided as text. The final user audio is
             the current query. [...]
  user       "Session timestamp: 2025-06-17 09:27"
  user       <audio: session 1, user turn 1>
  assistant  "The sentence with quotation marks is Mary said, 'I need to
             study harder.'"
  user       <audio: session 1, user turn 2>
  assistant  "The sentence with quotation marks is John said, 'he will meet
             us at seven o'clock.'"
  ...                                   (57 further turns across 8 sessions)
  user       "Session timestamp: 2025-09-01 07:44"
  user       <audio: final spoken query>
\end{lstlisting}

The instruction is identical for all models and budgets; its full text is in
Appendix~\ref{app:prompt_eval}. It states the audio/text role split, directs
the model to use information in the audio and not only the words spoken, warns
that user turns may come from different speakers, and allows abstention: if the
conversation does not determine one unique answer, the model should reply that
there is insufficient evidence. This clause is what makes AR items measurable,
and it is present in every evaluation run of Section~4; the acoustic-evidence
validation of \S\ref{sec:acoustic-validation}, which contains only answerable
items, uses the variant without it (Appendix~\ref{app:prompt_eval}).

\subsection{Reference Context-Length Measurement}
\label{app:length}

Audio and text are not commensurable in tokens, and every model tokenizes audio
differently, so a budget defined in any one model's units would make the length
axis model-dependent. We therefore measure length on one reference scale that
belongs to none of the evaluated systems: audio with a Whisper encoder at 50
tokens per second, and text with the Whisper tokenizer. The budget is a
property of the released item and is the same for every model; it does not say
how many tokens a given model spends on that item. Histories are built from
whole sessions and never by truncating one. The assembly procedure, the
length tolerance and the resulting history composition are given in
Appendix~\ref{app:assembly} and Figure~\ref{tab:dataset_stats}.

\subsection{Scoring and Judge Rubric}
\label{app:judge}

Responses are scored by Gemini-3.7-Flash with two rubrics, one for answerable
items and one for AR items, each returning a single JSON verdict of
\texttt{correct} or \texttt{incorrect}. The judge sees the final question, the
reference answer, the expected answer type and the model response. It never
sees the conversation history, so it cannot re-derive the answer and substitute
its own judgement for the annotation. For closed-set items the option list is
also supplied, so an off-list near-synonym is rejected on a closed-set item but
can be accepted on a free-text one.

For \textbf{answerable items}, the rubric asks whether the model's final
committed answer is semantically equivalent to the reference. It accepts
differences in case, punctuation, articles, number formatting and wording, as
well as numerically equivalent expressions; it requires every element of an
unordered set, and both the elements and their order for an ordered one. A
response is incorrect if it omits an answer-critical part, offers several
incompatible answers, contradicts its own final answer, or abstains.
Additional explanation is allowed only if it does not weaken the committed
answer.

For \textbf{AR items}, the rubric asks whether the model declined because the
evidence is insufficient. A response is correct when it states that the answer
cannot be determined because the conversation's evidence is insufficient,
missing or ambiguous; the exact phrase from the instruction is not required.
Three exclusions carry most of the weight. A bare ``I don't know'' is not
sufficient unless it attributes the uncertainty to the conversation. A response
that abstains and then guesses, or that offers several candidates, is
incorrect. A refusal on grounds of capability, policy or willingness (``I
cannot hear audio'', ``I have no access to previous conversations'') is not a
correct abstention, because it does not engage with the evidence, and crediting
it would reward the models that engage least. The judge is told to assume that
the annotation is correct and to evaluate only the response. Both judge prompts
are given in full in Appendix~\ref{app:prompt_eval}.

Overall accuracy pools answerable and AR items over all 799 questions. A failed
generation keeps its empty response and is scored incorrect, as is an item for
which the judge returned no parseable verdict. The only predictions removed
from a denominator are those a provider rejected before inference, since the
model never saw the history.

\subsection{Judge Cross-Validation}
\label{app:xjudge}

To test whether any conclusion depends on the judge, all 11{,}978 scored 8K
predictions of the fifteen models were re-scored by GPT-5.6-Luna with the same
prompt. The two judges agree on 97.95\% of predictions (Cohen's
$\kappa = 0.953$); the model ranking is unchanged (Spearman $\rho = 0.999$),
and the mean absolute per-model difference is 0.86~pp. Because the judge never
sees the history, its input does not depend on the budget; only the mix of
responses changes with length.

Because the production judge is a Gemini model and three evaluated systems are
Gemini models, we also test for family favoritism. Relative to GPT-5.6-Luna,
the production judge scores Gemini-family predictions 0.13~pp lower and all
other predictions 1.00~pp lower, so it favors Gemini outputs by 0.88~pp. This is
smaller than the gaps our conclusions rest on, but not smaller than every
pairwise gap: at 8K the three strongest models lie within 0.9~pp of each other
(Figure~\ref{fig:result}), so their order at that budget should not be read as meaningful.

\begin{table}[!tbp]
\centering\footnotesize
\begin{tabular}{@{}l r rr r rr@{}}
\toprule
Slice & $n$ & Gemini-3.7-Flash & GPT-5.6-Luna & $\Delta$ (pp) & Agreement & $\kappa$ \\
\midrule
All items & 11{,}978 & 31.1 & 31.9 & $+0.8$ & 98.0\% & 0.953 \\
\midrule
Speech Semantics & 3{,}478 & 48.0 & 48.6 & $+0.6$ & 98.7\% & 0.974 \\
Speaker Identity & 2{,}202 & 37.0 & 38.0 & $+1.0$ & 95.6\% & 0.907 \\
Paralinguistic Cues & 3{,}958 & 19.7 & 20.5 & $+0.8$ & 98.3\% & 0.948 \\
Environmental Sound & 2{,}340 & 19.7 & 20.7 & $+1.1$ & 98.4\% & 0.951 \\
\midrule
IE (answerable) & 3{,}447 & 26.3 & 26.4 & $+0.1$ & 97.5\% & 0.936 \\
MSR (answerable) & 3{,}313 & 35.0 & 35.7 & $+0.7$ & 99.0\% & 0.978 \\
TET (answerable) & 3{,}269 & 33.2 & 34.0 & $+0.8$ & 97.8\% & 0.950 \\
AR & 1{,}949 & 29.4 & 31.8 & $+2.4$ & 97.3\% & 0.936 \\
\bottomrule
\end{tabular}
\caption{\textbf{Cross-judge agreement at 8K}, over every scored prediction of
the fifteen models. The largest gap is on AR items: of the 50 AR items that the
second judge credits and the production judge rejects, 32 are capability
refusals (``I do not have access to previous conversations''), which the rubric
of Appendix~\ref{app:judge} excludes. The production judge applies that rule
and the second judge does not; where the two disagree, the reported scores
follow the stricter reading.}
\label{tab:c3}
\end{table}

\subsection{Models Included in the 64K Analysis}
\label{app:sixtyfour}

Ten of the fifteen systems accept a complete 64K history. They form the model
set wherever 64K results are reported, including Figure~\ref{fig:scaling},
where the same ten models are used at every budget. Table~\ref{tab:c4} marks
which systems completed the 64K run. Every model was attempted at 64K.

\begin{table}[!tbp]
\centering\footnotesize
\begin{tabular}{@{}l l@{}}
\toprule
\textbf{Model} & \textbf{64K Status} \\
\midrule
Gemini-3.1-Pro                 & \checkmark \\
Gemini-3.8-Flash               & \checkmark \\
Gemini-2.5-Pro                 & \checkmark \\
Qwen3-Omni-30B-A3B             & \checkmark \\
Gemma-4-E4B-it                 & \checkmark \\
Ultravox-v0.6-Llama-3.1-8B     & \checkmark \\
Audio-Flamingo-Next            & \checkmark \\
Baichuan-Audio-7B              & \checkmark \\
FireRedAudio-9B                & \checkmark \\
Baichuan-Omni-1.5-7B           & \checkmark \\
\midrule
Qwen3.8-Omni-Flash             & $\times$  \\
Qwen3.5-Omni-Plus              & $\times$  \\
MiniCPM-o-4.5                  & $\times$  \\
MiMo-Audio-7B-Instruct         & $\times$  \\
Phi-4-Multimodal               & $\times$ \\
\bottomrule
\end{tabular}
\caption{\textbf{The 64K model set.} \checkmark\ indicates full 64K evaluation completed for all 799 items.}
\label{tab:c4}
\end{table}


%

\section{Additional Results and Robustness}
\label{app:results}

All numbers in this appendix come from the same judged predictions as the main
text; no new inference was run. Results at 64K use the ten models of
Appendix~\ref{app:sixtyfour}.

\paragraph{Denominators.} Unless stated otherwise, each model is scored on the
items it was judged on at every budget it was run at, so the columns of one row
are computed on identical items and can be paired (\S\ref{app:paired}).
Figure~\ref{fig:opxev_tiers} instead uses one shared set, the ten 64K models
and the items judged at all four budgets, so its panels differ only in history
length; its values therefore differ from Figure~\ref{fig:op-by-type}, which averages
fifteen models at 32K.

\subsection{Full 64K Results}
\label{app:full64k}

\begin{table}[!tbp]
\centering\footnotesize
\begin{tabular}{@{}l rrrr r@{}}
\toprule
Model & Speech Sem. & Speaker & Paraling. & Environ. & Overall \\
\midrule
Gemini-3.1-Pro             & 68.5 & 26.5 & 19.7 & 15.4 & 34.3 \\
Gemini-3.8-Flash           & 50.9 & 38.1 & 16.7 & 15.4 & 30.3 \\
Ultravox-v0.6-Llama-3.1-8B & 28.0 & 29.3 & 12.9 & 17.9 & 21.3 \\
Gemini-2.5-Pro             & 35.8 & 14.3 & 14.4 & 16.7 & 21.0 \\
Qwen3-Omni-30B-A3B         & 31.5 & 21.8 & 15.9 & 11.5 & 20.7 \\
Gemma-4-E4B-it             & 27.2 & 24.5 & 12.9 & 11.5 & 18.9 \\
Audio-Flamingo-Next        & 26.3 & 22.4 & 12.9 &  9.0 & 17.8 \\
Baichuan-Audio-7B          & 26.7 & 29.3 &  9.8 &  7.1 & 17.8 \\
Baichuan-Omni-1.5-7B       & 20.3 & 13.6 & 11.4 & 11.5 & 14.4 \\
FireRedAudio-9B            & 22.0 & 11.6 & 10.2 &  8.3 & 13.5 \\
\midrule
Mean proprietary (3)       & 51.7 & 26.3 & 16.9 & 15.8 & 28.5 \\
Mean open-weight (7)       & 26.0 & 21.8 & 12.3 & 11.0 & 17.8 \\
Mean all (10)              & 33.7 & 23.1 & 13.7 & 12.4 & 21.0 \\
\bottomrule
\end{tabular}
\caption{\textbf{Accuracy (\%) at 64K} by answer-critical evidence type, for
the ten models of Appendix~\ref{app:sixtyfour}. Cells contain 232, 147, 264 and
156 questions (799 overall), including AR items as in
Figure~\ref{fig:result}.}
\label{tab:full64k}
\end{table}

Table~\ref{tab:full64k} completes Figure~\ref{fig:result} at 64K. The pattern
of the shorter budgets holds: speech semantics is highest, speaker identity
second, and paralinguistic cues and environmental sound lowest and close
together. The proprietary advantage is concentrated in speech semantics (51.7
against 26.0); on the three audio-native types the two groups are within five
points of each other.

\subsection{Operation $\times$ Evidence Across History Lengths}
\label{app:opxev}

\begin{figure}[!tbp]
\centering
\includegraphics[width=\linewidth]{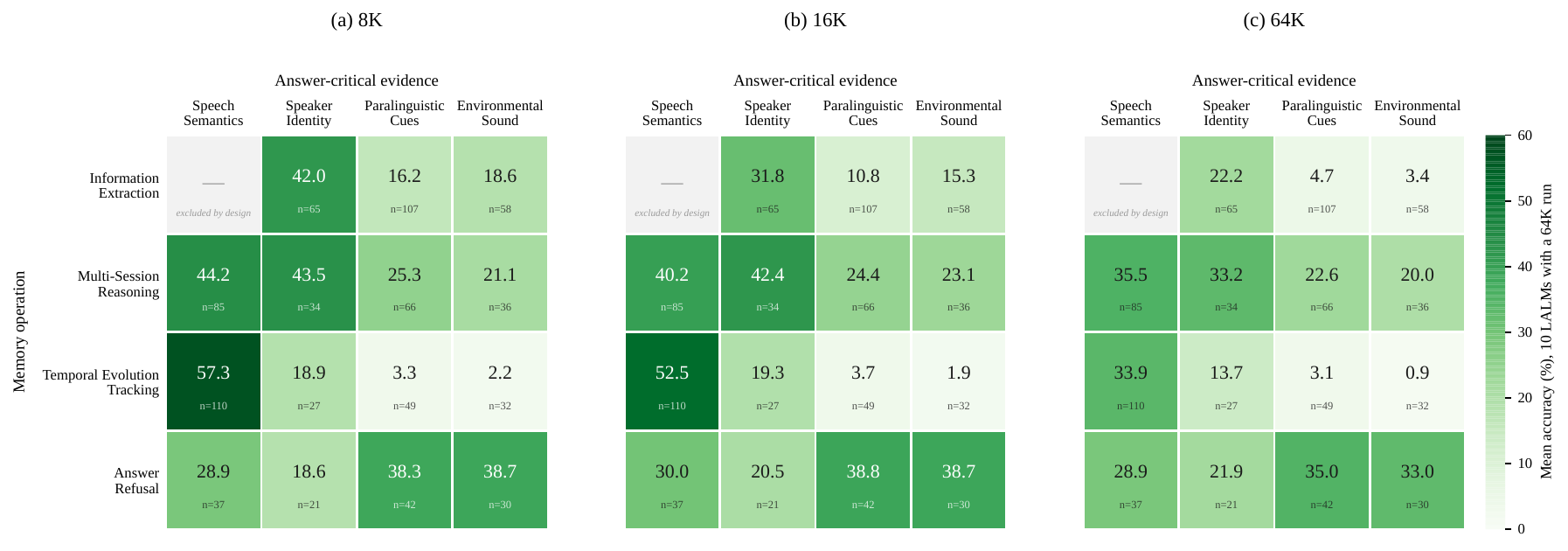}
\caption{\textbf{Mean accuracy (\%) for each operation $\times$ evidence cell at
8K, 16K and 64K}, over the ten 64K models and the items judged at all four
budgets. The 32K values over all fifteen models are in
Figure~\ref{fig:op-by-type}. Speech-Semantics IE is excluded by design.}
\label{fig:opxev_tiers}
\end{figure}

The interaction of Figure~\ref{fig:op-by-type} holds at every length
(Figure~\ref{fig:opxev_tiers}). Temporal Evolution Tracking on speech semantics
is the strongest cell at 8K and 16K (57.3 and 52.5) and remains among the
strongest at 64K (33.9, behind semantic Multi-Session Reasoning at 35.5 and
paralinguistic Answer Refusal at 35.0), while the same operation on acoustic states is the weakest (3.3, 3.7 and
3.1 on paralinguistic cues; 2.2, 1.9 and 0.9 on environmental sound).
Among the answerable operations, Multi-Session Reasoning varies least across
the three audio-native evidence types, and the
Answer Refusal row is highest on paralinguistic and environmental evidence, at
every length. Longer histories lower the cells but do not change this pattern.

\subsection{Answerable versus Answer Refusal Performance}
\label{app:ar_results}

\begin{table}[!tbp]
\centering\footnotesize
\begin{tabular}{@{}l rr@{}}
\toprule
Model & Answerable ($n{=}669$) & AR ($n{=}130$) \\
\midrule
Gemini-3.1-Pro             & 29.0 & 61.5 \\
Gemini-3.8-Flash           & 26.6 & 49.2 \\
Ultravox-v0.6-Llama-3.1-8B & 21.1 & 22.3 \\
Audio-Flamingo-Next        & 20.6 &  3.1 \\
Qwen3-Omni-30B-A3B         & 20.2 & 23.1 \\
Baichuan-Audio-7B          & 20.2 &  5.4 \\
Gemma-4-E4B-it             & 17.9 & 23.8 \\
FireRedAudio-9B            & 15.1 &  5.4 \\
Gemini-2.5-Pro             & 12.0 & 67.7 \\
Baichuan-Omni-1.5-7B       &  8.4 & 45.4 \\
\bottomrule
\end{tabular}
\caption{\textbf{Accuracy (\%) at 64K on answerable and AR items.} AR accuracy
is the rate at which the model correctly declined. The overall scores of
Table~\ref{tab:full64k} are the item-weighted mean of the two columns.}
\label{tab:ar_split}
\end{table}

The overall score of the main text pools answerable and AR items;
Table~\ref{tab:ar_split} separates them. The two columns rank the models almost
independently (Spearman $\rho = -0.02$). Gemini-2.5-Pro is second to last on
answerable items and first on AR items, and more than half of its overall score
comes from AR items; Audio-Flamingo-Next is mid-table on answerable items and
near zero on AR items.

This separation is why the overall score pools the two. Answering from the
history and recognizing that the history does not support an answer are
distinct requirements of conversational memory, and the near-zero correlation
shows that neither column alone measures both. Pooling also bounds what either
strategy achieves on its own: a system that always abstains would score 16.3\%
overall (130 of 799 items) and 0\% on every answerable item, while a system
that never abstains forgoes all AR items. The two models at the top of the
overall ranking, Gemini-3.1-Pro and Gemini-3.8-Flash, are also the two
strongest on answerable items, so the leading positions are not obtained
through refusal. Both columns are reported so that either can be used on its
own.

\begin{table}[!tbp]
\centering\footnotesize
\begin{tabular}{@{}l l rrrr c rrr@{}}
\toprule
& & \multicolumn{4}{c}{By evidence type} & & \multicolumn{3}{c}{By source operation} \\
\cmidrule(lr){3-6}\cmidrule(l){8-10}
Budget & Models & Sem. & Spk. & Para. & Env. & & IE & MSR & TET \\
\midrule
8K  & 15 & 22.0 & 16.2 & 35.2 & 39.3 & & 27.2 & 32.9 & 27.3 \\
16K & 15 & 23.6 & 14.6 & 35.6 & 37.6 & & 28.8 & 31.7 & 26.5 \\
32K & 15 & 19.9 & 16.2 & 34.3 & 34.2 & & 28.0 & 28.2 & 25.0 \\
64K & 10 & 28.9 & 21.9 & 35.0 & 33.0 & & 31.9 & 32.9 & 25.9 \\
\midrule
\multicolumn{2}{@{}l}{AR items} & 37 & 21 & 42 & 30 & & 47 & 49 & 34 \\
\bottomrule
\end{tabular}
\caption{\textbf{Mean AR accuracy (\%)} by evidence type and by the operation of
the source item. The last row gives the number of AR items per column; the
Speaker column rests on 21 items.}
\label{tab:ar_breakdown}
\end{table}

By evidence type (Table~\ref{tab:ar_breakdown}), AR accuracy is highest on
paralinguistic and environmental items, the two types with the lowest
answerable accuracy, and lowest on speaker identity, at every budget. By source
operation the differences are small. The variation in AR accuracy therefore
follows the evidence type rather than the operation.

\subsection{Paired Uncertainty for Context Scaling}
\label{app:paired}

The same question and evidence appear at every budget, so the 8K and 64K scores
of one model are paired at the item level. Intervals are percentile 95\%
bootstrap intervals over items (10{,}000 resamples), with both budgets rescored
on each resample.

\begin{table}[!tbp]
\centering\footnotesize
\begin{tabular}{@{}l r rr ll@{}}
\toprule
Evidence type & $n$ & 8K & 64K & Drop (pp) [95\% CI] & Retention [95\% CI] \\
\midrule
Speech Semantics    & 232 & 48.0 & 33.7 & 14.3 [11.9, 16.7] & 0.703 [0.666, 0.741] \\
Speaker Identity    & 147 & 34.8 & 23.1 & 11.6 [8.0, 15.4]  & 0.665 [0.590, 0.749] \\
Paralinguistic Cues & 264 & 19.6 & 13.7 & 5.9 [4.1, 7.8]    & 0.698 [0.626, 0.775] \\
Environmental Sound & 156 & 19.7 & 12.4 & 7.2 [4.1, 10.7]   & 0.632 [0.511, 0.764] \\
\midrule
Overall             & 799 & 30.6 & 21.0 & 9.6 [8.3, 11.0]   & 0.685 [0.653, 0.718] \\
\bottomrule
\end{tabular}

\vspace{6pt}
\begin{tabular}{@{}l rr ll@{}}
\toprule
Model & 8K & 64K & Drop (pp) [95\% CI] & Retention [95\% CI] \\
\midrule
Gemini-3.1-Pro             & 42.7 & 34.3 & 8.4 [5.4, 11.4]   & 0.804 [0.741, 0.870] \\
Gemini-3.8-Flash           & 41.9 & 30.3 & 11.6 [8.4, 15.0]  & 0.722 [0.655, 0.794] \\
Gemini-2.5-Pro             & 32.3 & 21.0 & 11.3 [8.0, 14.5]  & 0.651 [0.573, 0.737] \\
Qwen3-Omni-30B-A3B         & 32.3 & 20.7 & 11.6 [8.5, 14.8]  & 0.640 [0.562, 0.726] \\
Gemma-4-E4B-it             & 31.3 & 18.9 & 12.4 [9.0, 15.8]  & 0.604 [0.521, 0.693] \\
Ultravox-v0.6-Llama-3.1-8B & 28.5 & 21.3 & 7.3 [4.1, 10.5]   & 0.746 [0.650, 0.848] \\
Audio-Flamingo-Next        & 27.2 & 17.8 & 9.4 [6.8, 12.0]   & 0.654 [0.575, 0.737] \\
Baichuan-Audio-7B          & 26.2 & 17.8 & 8.4 [5.5, 11.3]   & 0.679 [0.591, 0.777] \\
FireRedAudio-9B            & 24.2 & 13.5 & 10.6 [7.9, 13.5]  & 0.560 [0.472, 0.656] \\
Baichuan-Omni-1.5-7B       & 19.9 & 14.4 & 5.5 [2.3, 8.8]    & 0.723 [0.592, 0.873] \\
\bottomrule
\end{tabular}
\caption{\textbf{Paired change from 8K to 64K}, as the mean over the ten 64K
models by evidence type (top) and per model pooled over evidence types
(bottom). Retention is 64K accuracy divided by 8K accuracy.}
\label{tab:paired}
\end{table}

Every interval on the drop in Table~\ref{tab:paired} excludes zero: the fall
from 8K to 64K is supported by the item sample for every model and every
evidence type. The retention estimates order the evidence types as in
Figure~\ref{fig:scaling}: speech semantics (0.703) and paralinguistic cues
(0.698) retain the most, speaker identity (0.665) and environmental sound
(0.632) the least. The absolute drops differ more (14.3
pp on speech semantics against 5.9 pp on paralinguistic cues), mainly because
the types start from different accuracies at 8K.

\subsection{Baseline Difficulty versus Context Sensitivity}
\label{app:baseline}

\begin{figure}[!tbp]
\centering
\includegraphics[width=\linewidth]{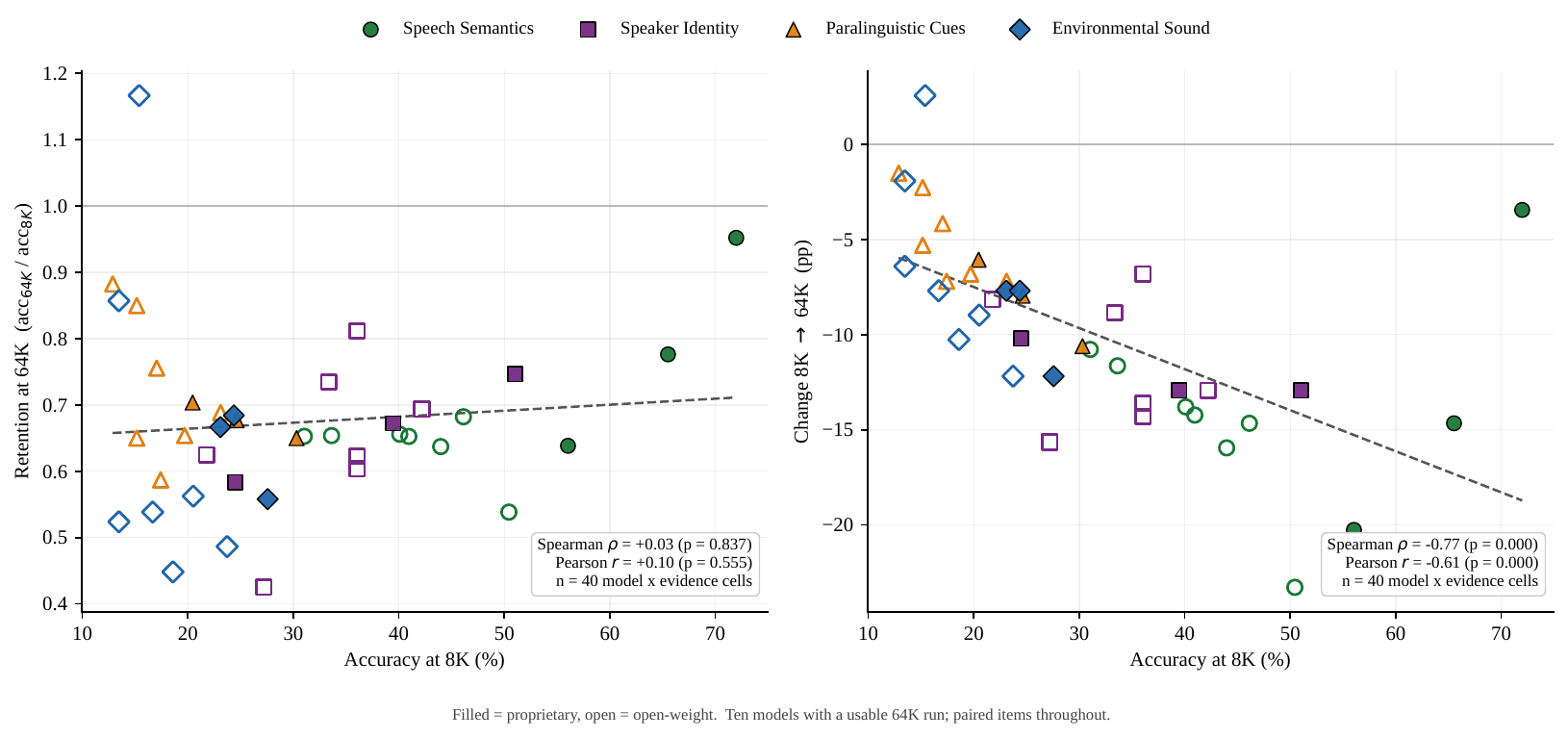}
\caption{\textbf{Baseline accuracy against sensitivity to length}, one point
per model $\times$ evidence type (40 points). Left: retention at 64K against 8K
accuracy. Right: change in percentage points against 8K accuracy.}
\label{fig:baseline}
\end{figure}

At the level of model $\times$ evidence cells, retention shows no detectable
relation to 8K accuracy (Spearman $\rho = 0.03$, $p = 0.84$), whereas the
change in percentage points correlates strongly with it ($\rho = -0.77$,
$p < 0.001$). The second correlation is largely produced by the
floor: a cell at 5\% accuracy at 8K cannot lose twenty points, while its
retention is not bounded in the same way. This supports, at 40 points, the
main-text observation that low absolute accuracy and sensitivity to additional
history are different failure modes; retention is the measure that separates
them.
\section{Error Analysis Details}
%


The error analysis considers only incorrect responses that commit to an
answer: incorrect refusals, empty generations and responses that stop
mid-sentence are excluded before attribution (see \emph{Exclusions} below),
and refusals are analyzed separately in Appendix~\ref{app:ar_results}. All
proportions in Figure~\ref{fig:errors} and this appendix are macro-averaged
over model $\times$ evidence-type cells with at least five errors, so they
differ from the pooled counts of \S\ref{app:err_counts}.
Figure~\ref{fig:errors} divides the remaining incorrect answers into five
attribution classes. This appendix gives the class definitions
(\S\ref{app:err_taxonomy}), the attribution procedure
(\S\ref{app:err_procedure}), the breakdown by question subtype
(\S\ref{app:err_subtype}), the counts behind every proportion
(\S\ref{app:err_counts}) and representative cases (\S\ref{app:err_cases}).

\paragraph{Population.} All results use the 64K budget, the three audio-native
evidence types (Speaker Identity, Paralinguistic Cues, Environmental Sound) and
the ten systems with a complete 64K run (Appendix~\ref{app:sixtyfour}): 474
answerable items $\times$ 10 systems $=$ 4{,}740 attempts, of which 4{,}137 are
scored incorrect.

\paragraph{Exclusions.} Three kinds of incorrect response contain no committed
answer to trace and are removed before attribution: refusals, whether
evidence-based or capability-based (1{,}595); empty generations (167); and
answers that stop mid-sentence (162). Refusals are analyzed separately in
Appendix~\ref{app:ar_results}. The remaining 2{,}213 incorrect answers are all
attributed.

\subsection{Error Attribution Taxonomy}
\label{app:err_taxonomy}

Table~\ref{tab:err_taxonomy} defines the five classes. Each class is derived
by fixed rules from the audit record of \S\ref{app:err_procedure}, never from
the response alone.

\begin{table}[!tbp]
\centering\footnotesize
\begin{tabular}{@{}p{0.19\textwidth}p{0.62\textwidth}p{0.11\textwidth}@{}}
\toprule
Class & Definition & Example \\
\midrule
Evidence recognition / localization
  & The answer-critical evidence was not recovered. The response commits to
    nothing, to a generic description or to an answer of the wrong type
    without identifying the cue; or it commits to content from a non-evidence
    session on a question whose retrieval key is not a cue; or it reads a
    wrong value from an evidence session.
  & Cases 1, 2 \\
\addlinespace[2pt]
Binding / association
  & Real content from the conversation is attached to the wrong speaker,
    session or fact: the audit places it with another speaker, or it comes
    from a non-evidence session on a question whose retrieval key is a cue
    (Cue-to-Fact, Matching).
  & Case 3 \\
\addlinespace[2pt]
Operation execution
  & The evidence was located but the operation over it failed: a wrong count,
    another point on the same state trajectory, or a wrongly decided
    comparison; or the cue is identified but the requested operation is not
    carried out.
  & Cases 5, 5b \\
\addlinespace[2pt]
Unsupported answer
  & The response commits to a specific answer of the right type that no
    session supports, including invented inventories of sounds or speakers;
    also assigned to any committed answer that no stage rule covers.
  & Case 6 \\
\addlinespace[2pt]
Not attributable
  & The audit judges the committed answer equivalent to the reference.
    Reported as its own class, never redistributed.
  & --- \\
\bottomrule
\end{tabular}
\caption{\textbf{Error attribution classes.} Cases refer to
\S\ref{app:err_cases}.}
\label{tab:err_taxonomy}
\end{table}

The classes follow the order in which a memory query can fail: the evidence is
recognized, then bound to the right speaker or session, then operated on. Each
error is assigned the earliest stage at which it fails. \emph{Unsupported
answer} is defined by the answer's relation to the history rather than by a
stage, and is assigned only when no stage rule applies.

\subsection{Attribution Procedure}
\label{app:err_procedure}

\paragraph{Inputs.} Each incorrect response is audited by Gemini-3.7-Flash, which
receives four inputs.
\begin{enumerate}
\item \textbf{The full conversation as text}, in session order. Each session is
  marked \texttt{[EVIDENCE]}, \texttt{[DISTRACTOR]} or \texttt{[FILLER]}
  according to the item's evidence chain, and filler sessions are shortened to
  a one-line summary. User turns are verbatim transcripts with the three
  acoustic channels written in, because the evaluated model heard them and a
  text reader cannot: inline emotion, prosody and style markers; one line
  naming the background sound of the session; and speaker labels numbered by
  first appearance.
\item \textbf{The question}, as the evaluated system received it.
\item \textbf{The reference answer.}
\item \textbf{The model response.}
\end{enumerate}

\paragraph{The auditor does not choose a class.} It is never shown the
taxonomy. It answers factual questions about where the answer came from and
returns a structured record: whether the response commits to an answer, whether
it identified the answer-critical cue, which session the answer came from, and
how the answer relates to the reference. The class is then derived from this
record in code by a fixed, ordered rule set released with the code. The auditor
therefore cannot anchor on a class name, and every assignment can be checked
against the line quoted in \texttt{why} without re-running anything. The full
prompt is given in Appendix~\ref{app:prompt_attr}.

\paragraph{Ambiguous cases.} Three cases are fixed by rule rather than left to
judgement. (i) A response that names the correct cue but does not perform the
requested operation (``I heard a crackling fire'' to a question asking how many
times) is Operation execution, not Recognition; the audit records whether the cue was
identified, which separates the two. (ii) A same-or-different
verdict is computed over two sessions rather than retrieved from one, so a
verdict decided wrongly after both sessions were located is Operation
execution. (iii) 45 errors (2.0\%) are Not attributable and are reported as
their own class rather than redistributed.

\subsection{Breakdown by Question Subtype}
\label{app:err_subtype}

\begin{figure}[!tbp]
\centering
\includegraphics[width=\linewidth]{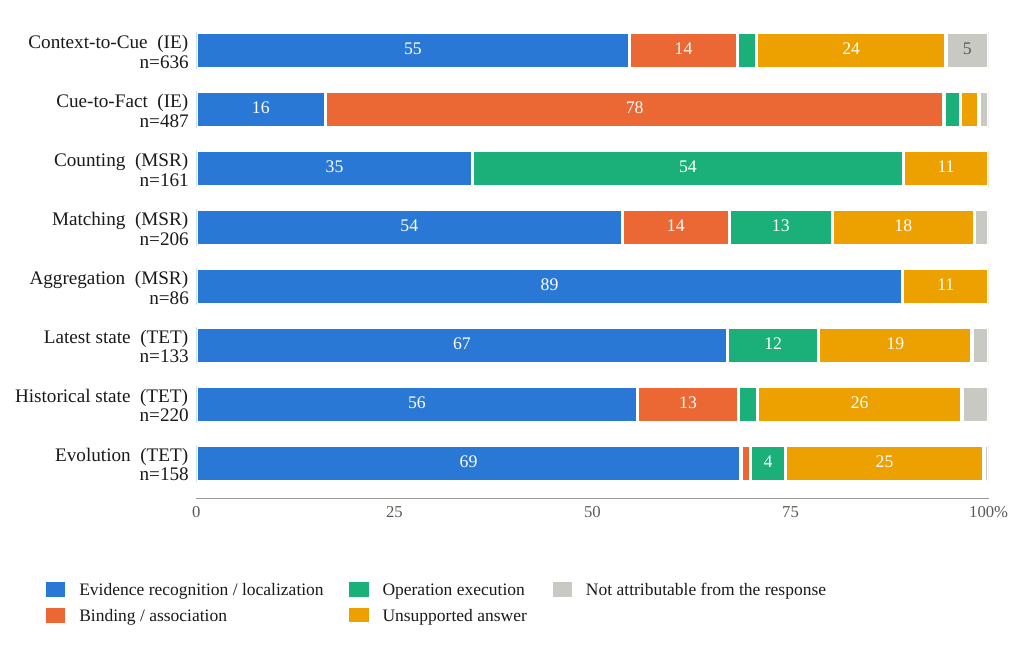}
\caption{\textbf{Error composition by question subtype at 64K}, in the five
classes of Figure~\ref{fig:errors}, macro-averaged over model $\times$
evidence-type cells with at least five errors. Counts are in
Table~\ref{tab:err_counts}.}
\label{fig:err_subtype}
\end{figure}

Figure~\ref{fig:err_subtype} breaks Figure~\ref{fig:errors} down by the eight
answerable subtypes of Table~\ref{tab:operation_subtypes}. As in
Figure~\ref{fig:errors}, proportions are macro-averaged over model $\times$
evidence-type cells with at least five errors. 

Three patterns are visible only at this level.
\begin{itemize}
\item \textbf{Binding errors concentrate in Cue-to-Fact.} 78\% of Cue-to-Fact
  errors are binding failures, against 0--14\% in every other subtype.
  Cue-to-Fact is the subtype in which the cue is the retrieval key and a fact
  is the answer, so the model must attach what it retrieves to the session the
  cue identifies.
\item \textbf{Operation errors concentrate in Counting.} They account for 54\%
  of Counting errors, 13\% of Matching errors, at most 12\% elsewhere and 1\% (1
  of 101) in Aggregation. Only 4 errors in the whole population report a real value of
  the tracked attribute at the wrong point in time; most temporal errors occur
  before the state sequence is assembled.
\item \textbf{Aggregation errors are almost all recognition failures} (89\%).
  Asked to compare two groups of cues, responses compare the topics of the two
  threads rather than the cues.
\end{itemize}

\subsection{Absolute Counts}
\label{app:err_counts}

Table~\ref{tab:err_counts} gives the unweighted counts behind
Figure~\ref{fig:err_subtype}. They differ from the macro-averaged proportions
where a subtype's errors are unevenly spread across systems. ``Cells'' is the
number of model $\times$ evidence-type cells with at least five errors, i.e.\
the units the macro-average is taken over.

\begin{table}[!tbp]
\centering\footnotesize
\begin{tabular}{@{}ll rrrrr rr@{}}
\toprule
Op. & Subtype & Recog. & Bind. & Oper. & Unsup. & N.a. & Total & Cells \\
\midrule
IE & Context-to-Cue & 428 & 58 & 16 & 116 & 23 & 641 & 26 \\
IE & Cue-to-Fact & 85 & 387 & 12 & 9 & 6 & 499 & 26 \\
MSR & Counting & 61 & 0 & 97 & 18 & 1 & 177 & 20 \\
MSR & Matching & 132 & 19 & 31 & 37 & 7 & 226 & 19 \\
MSR & Aggregation & 87 & 0 & 1 & 12 & 1 & 101 & 11 \\
TET & Latest state & 99 & 0 & 17 & 29 & 2 & 147 & 14 \\
TET & Historical state & 152 & 27 & 3 & 50 & 4 & 236 & 20 \\
TET & Evolution & 117 & 4 & 8 & 56 & 1 & 186 & 17 \\
\midrule
\multicolumn{2}{@{}l}{Speaker Identity} & 163 & 244 & 82 & 45 & 23 & 557 & \\
\multicolumn{2}{@{}l}{Paralinguistic Cues} & 707 & 178 & 71 & 49 & 18 & 1{,}023 & \\
\multicolumn{2}{@{}l}{Environmental Sound} & 291 & 73 & 32 & 233 & 4 & 633 & \\
\midrule
\multicolumn{2}{@{}l}{Total} & 1{,}161 & 495 & 185 & 327 & 45 & 2{,}213 & \\
\bottomrule
\end{tabular}
\caption{\textbf{Attributed errors at 64K}, by subtype (top) and by
evidence type (middle). Aggregation (11 cells) and Latest state (14 cells)
rest on the fewest cells, and their proportions should be read with that in
mind; every subtype has at least 101 attributed errors.}
\label{tab:err_counts}
\end{table}

\subsection{Representative Failure Cases}
\label{app:err_cases}

Six cases, each a real (item, system) pair; the quoted evidence is the line
the audit cited.

\paragraph{Case 1: Recognition, the cue is not identified.} Item 001307,
Ultravox, Aggregation / Paralinguistic. Reference: \emph{different}. Question:
``How do the two vocal delivery groups compare?'' Response: ``\ldots the two
vocal delivery groups are similar in their emphasis on prioritizing the produce
pickup, but the grocery errands thread also mentions the importance of\ldots''
The response compares what the two threads are about; no delivery is named.

\paragraph{Case 2: Localization, content outside the evidence chain.} Item
003248, Audio-Flamingo-Next, Counting / Speaker. Reference: \emph{2}.
Response: ``There are 10 chats that fit the specified speaker identity
condition.'' Ten sessions mention the report review; two carry the speaker
condition the question imposes. The system counted the topic, not the
condition.

\paragraph{Case 3: Binding, right content, wrong session.} Item 000005,
Gemini-2.5-Pro, Cue-to-Fact / Paralinguistic. Reference: \emph{lower drawer}.
Response: ``Probably that storage box is hanging inside the hanging caddy.''
The line is verbatim from Session 26, but the question's delivery cue points to
a different session; the system matched on topic.

\paragraph{Case 5: Operation, members located, count wrong.} Item 000128,
Gemini-3.8-Flash, Counting / Speaker. Reference: \emph{3}. Response: ``2'',
against three qualifying mentions in Sessions 16, 31 and 52.

\paragraph{Case 5b: Operation, both sides located, comparison wrong.} Item
000163, Qwen3-Omni-30B-A3B, Matching / Paralinguistic. Reference:
\emph{false}. Response: ``Yes, both chats about the hallway cabinet are linked
by the same vocal delivery.'' Session 11 is rendered reflective and slow,
Session 46 sad and slow. 
Both deliveries were described; the comparison came out wrong.

\paragraph{Case 6: Unsupported, the inventory is invented.} Item 000738,
Gemini-3.8-Flash, Aggregation / Environmental. Reference: \emph{different}.
Response: ``There are 3 sessions with background noises (kitchen
cleaning/prep sounds) and 3 sessions without background noises (quiet
speech). Thus, both sound groups have an equal count of 3.'' The evidence
sessions carry \emph{engine} and \emph{hen}; neither the categories nor the
counts exist in the conversation.

\medskip
\noindent Cases 1 and 6 are the two ways the acoustic channel goes unused:
describing it generically or inventing its content. Case 3 is a binding
failure: the right content attached to the wrong session. Cases 5 and 5b are
Operation execution, the only class in which the evidence was located and
correctly bound but then mishandled; it accounts for 8.4\% (185) of attributed errors.



%

\section{Prompts}
\label{app:prompts}

Prompts are reproduced from the artifacts that produced the released data;
runtime placeholders appear in the form the assembler substitutes, and
passages marked \texttt{[...]} are abridged. This appendix gives the prompts
that determine what the benchmark contains and how it is scored. All prompts,
in full, are released with the code.

\subsection{Question and Evidence Generation Prompt}
\label{app:prompt_qgen}

One system prompt is shared by the three answerable operations; the user prompt
is selected by operation and carries the generation request and the four
component schemas as runtime placeholders. AR items use no generation prompt:
they are derived from admitted answerable items (Appendix~\ref{app:ar_construction}).

\begin{promptbox}[colback=black!5, colframe=white!40!black, title=Question and Evidence Generation Prompt (shared system prompt)]{}
\scriptsize
\begin{verbatim}
[SYSTEM]

You design one structured candidate for an English benchmark of long-term memory
across multiple spoken-dialogue sessions.

Your job is to specify the question logic and the answer-critical evidence. You are
not writing the final spoken query or any conversation turns.

SCOPE

Generate exactly one candidate bundle containing:
1. question_spec
2. operation_extension
3. atomic_evidence
4. evidence_type_extension

Do not generate:
- user-assistant dialogue;
- final spoken query wording;
- assistant replies;
- text-to-speech instructions;
- environmental mixing instructions;
- haystack or filler sessions;
- assembled histories;
- an answer-refusal item;
- explanations outside the required JSON.

GLOBAL CONTRACT

1. Use English for every machine-readable string.
2. Use exactly one answer-critical evidence type.
3. The gold answer must be uniquely determined by the required evidence.
4. The requested memory operation must be necessary to derive the answer.
5. Every required evidence item must be individually necessary. Removing one
   required item must make the item unanswerable, ambiguous, or change the correct
   answer.
6. Evidence necessity is operation-specific:
   - IE: every declared evidence item must be necessary.
   - MSR: every declared evidence operand must be individually necessary
     (leave-one-out).
   - TET historical_state / evolution_transition: necessity follows the
     subtype-specific temporal contract; every required node must contribute a
     necessary state or transition.
   - TET latest_state: superseded states MAY be marked supportive temporal-chain
     evidence (necessity="supportive", removal_effect="none"); the terminal decisive
     valid update MUST be marked answer-required (necessity="required",
     removal_effect="answer_changed"). At least one node must be marked required.
   Outside these allowances, do not include supportive-only evidence in the required
   evidence set.
7. The answer must not be recoverable from the question alone, common knowledge, or
   an unstated real-world assumption.
8. For speaker, paralinguistic, and environmental routes, transcript semantics alone
   must not reveal the answer-critical cue.
9. Do not name or paraphrase a non-semantic cue in text that will later be realized
   as dialogue.
10. Do not make age, gender, accent, culture, or another demographic inference part
    of the answer.
11. The evidence must be realizable later as ordinary spoken English. Each semantic
    fact should fit naturally into one or two spoken sentences.
12. The assistant must be able to reply coherently without repeating, confirming, or
    naming the answer-bearing fact or acoustic cue.
13. Keep different future session slots locally self-contained unless the requested
    operation explicitly requires a cross-session relation.
14. Do not create an alternative answer route through redundant facts, aliases,
    timestamps, ownership, topic choice, or descriptive wording.
15. Do not generate an answer-refusal candidate. Answer-refusal items are derived
    later by changing evidence sufficiency while preserving the query.
16. Use only fact, evidence, and session-slot identifiers permitted by
    generation_request.id_policy. Artifact-level IDs in the request are runner
    metadata and must not be inserted into schemas that do not define an ID field.
17. The `id_policy.session_slot_id_prefix` is reserved for downstream Session
    Realization and MUST NOT be used for planning-stage evidence identifiers. Every
    `evidence_type_extension.items[].evidence_id` and every Operation Extension
    reference to an evidence item MUST use `id_policy.evidence_id_prefix`.
18. This includes the IE answer-bearing evidence references, MSR
    `spec.evidence_set`, and TET `spec.timeline[].evidence_id`.
19. Follow the injected JSON Schemas exactly. Do not add fields that are not allowed
    by those schemas.
20. Every enum-valued field accepts ONLY the values listed in that field's own enum
    in the injected schema. Do not reuse a value that belongs to a different,
    similarly-named, or neighbouring field. In particular:
   - question_spec.selection.condition_type does NOT accept "speaker_identity" (that
     value belongs to operation_extension matching_spec.resolution_key); use
     "query_speaker_identity" or "speaker_relation".
   - atomic_evidence facts[].reasoning_roles does NOT accept "required" or
     "supportive" (those belong to answer_contribution.necessity).
   - question_spec.validity_constraints.cue_ablation_expected does NOT accept
     "answer_changed" (that belongs to answer_contribution.removal_effect).
21. Treat generation_request.content_blueprint as a binding production contract, not
    as optional inspiration.
22. Copy content_blueprint.required_operation_extension exactly into
    operation_extension. Do not change its operator, references, grouping, matching,
    timeline, state values, temporal query, or transition output.
23. Copy content_blueprint.required_answer_spec and
    content_blueprint.required_validity_constraints exactly into the corresponding
    question_spec fields.
24. Create one evidence item for every entry in
    content_blueprint.evidence_payload_plan, using the exact evidence_id, payload
    field, and payload value specified there. For speaker evidence, also use
    content_blueprint.required_query_speaker_id exactly.
25. Realize every entry in content_blueprint.atomic_fact_plan as a bound Atomic
    Evidence fact with the specified fact_id, entity_id, predicate, object value,
    reasoning roles, necessity, and removal effect.
26. The semantic plan controls topic, target entity, target attribute, and session
    anchor. Preserve its meaning while writing concise schema-compatible strings.
27. Never expose blueprint_id, blueprint_fingerprint, control-field names, or
    forbidden_surface_mentions in transcript-realizable semantic content.
28. Do not substitute synonyms for evidence payload labels, speaker slots, operation
    enums, identifiers, or gold values.

EVIDENCE-TYPE CONTRACTS

speech_semantics:
- The answer-critical information is carried by spoken semantic content.
- Speaker identity, prosody, delivery style, and environmental sound are
  non-informative.
- This evidence type is allowed for MSR and TET, but not IE.

speaker:
- The answer depends on binding a fact, participation record, or history segment to
  the correct speaker.
- Canonical speaker identity must not be visible in transcript text.
- Use abstract speaker slots supplied by the generation request.

paralinguistic:
- The answer depends on one discrete cue from the approved paralinguistic inventory.
- The cue must not be named or paraphrased in transcript-realizable text.
- Do not perform arithmetic over continuous emotion intensity.

environmental:
- The answer depends on an audible environmental event or acoustic scene from the
  approved asset catalog.
- The sound must not be named or paraphrased in transcript-realizable text.
- Treat the evidence as an observed acoustic event. Do not infer a physical location
  unless spoken semantics independently state it.

OUTPUT CONTRACT

Return exactly one valid JSON object with this top-level structure:

{
  "question_spec": {
    "schema_version": "0.1",
    "question_spec": { ... }
  },
  "operation_extension": { ... },
  "atomic_evidence": {
    "schema_version": "0.1",
    "atomic_evidence": { "facts": [ ... ] }
  },
  "evidence_type_extension": { ... }
}

The value of question_spec is the complete Common Question Spec artifact.
The value of atomic_evidence is the complete Atomic Evidence artifact.
The other two values are the roots accepted by their schemas.
When generation_request.route.evidence_type is speech_semantics, set
evidence_type_extension.evidence_type to semantic.
The four values must conform exactly to the runtime schemas supplied in the user
message.

Do not return Markdown.
Do not return comments.
Do not return a rationale, analysis, self-critique, or chain of thought.
Do not wrap the JSON in a code fence.

Before returning, silently verify:
- route compatibility;
- answer uniqueness;
- evidence necessity;
- operation necessity;
- evidence-type necessity;
- absence of transcript-only and common-knowledge shortcuts;
- cross-reference consistency across the four output objects.
\end{verbatim}
\end{promptbox}

\subsection{Dialogue Generation Prompts}
\label{app:prompt_dialogue}

User turns and assistant replies are generated by separate calls
(Appendix~\ref{app:dialogue}). The user generator receives the session-local
plan and writes every user turn of one session; the assistant generator
receives only the visible conversation prefix and writes one reply.

\begin{promptbox}[colback=black!5, colframe=white!40!black, title=Dialogue Generation Prompt: User Turns (evidence-aware)]{}
\scriptsize
\begin{verbatim}
[SYSTEM]

You generate every USER spoken-text turn for one planned evidence session in an
English benchmark of long-term memory across multiple spoken-dialogue sessions.

This is a text-realization task. Do not synthesize audio.

ONE CALL, ONE SESSION

1. Generate all USER turns assigned to the current session in one response.
2. Generate no ASSISTANT replies.
3. Preserve the assigned USER turn IDs, order, count, and target word-count ranges.
4. Use the complete session-local plan to make the USER turns coherent as one
   conversation.
5. Treat each assistant_reply_contract only as an abstract expectation about the
   unseen reply between two USER turns. Do not write or quote the reply.

INPUT VISIBILITY

6. Use only the supplied session-local job.
7. Do not infer the final query, gold answer, full question specification, full
   operation extension, other sessions, haystack, filler, AR variant, or final
   history.
8. Do not write text that sounds engineered to answer a future benchmark question.
9. Do not mention internal IDs, facts, evidence items, routes, slots, labels,
   schemas, prompts, or benchmark metadata.

FACT REALIZATION

10. Realize every required local fact exactly once.
11. Realize each fact only in the USER turn to which it is bound.
12. Preserve subject, predicate, object value, polarity, epistemic status, temporal
    meaning, and reference policy.
13. Use natural first-person spoken English when required by the fact reference
    policy.
14. Embed the fact incidentally in the ongoing topic rather than presenting a
    database record.
15. Do not repeat the fact value in another USER turn.
16. Do not add another personal fact that changes, competes with, negates, or
    resolves the benchmark answer.
17. Supporting context may improve naturalness but must remain answer-neutral.
18. Do not copy exact wording from an imagined final query.

EVIDENCE-TYPE RULES

19. Speech Semantics:
    - the required semantic content must be recoverable from the transcript;
    - align the evidence to the exact semantic span.

20. Speaker:
    - do not name or describe the internal speaker identity;
    - do not include a self-introduction or demographic description to expose
      identity;
    - align the evidence to the whole audio turn;
    - the transcript must remain compatible with later voice assignment.

21. Paralinguistic:
    - do not name the cue label or describe how the user sounds;
    - do not write statements such as "I sound hesitant" or "I am whispering";
    - make the transcript naturally speakable in the requested style without making
      the style uniquely inferable from text alone;
    - align the evidence to the whole audio turn unless the input explicitly
      requires a local audio span.

22. Environmental:
    - do not mention the target sound or scene;
    - do not say that the user or assistant can hear it;
    - do not infer or state a physical location from the sound;
    - make the transcript semantically compatible with later mixing;
    - align the evidence to the whole audio turn.

CONVERSATIONAL QUALITY

23. Write natural, concise, conversational English suitable for speech.
24. Keep one coherent local topic across the session.
25. Let information unfold progressively rather than repeating the opening.
26. Use ordinary pronouns, contractions, and discourse connectives where natural.
27. Avoid formal reports, lists, headings, stage directions, narration, and
    quotation marks around the user's own speech.
28. Do not assume exact wording from unseen ASSISTANT replies.
29. After a follow_up or clarify contract, write the next USER turn so it can
    naturally follow a broad relevant reply, not one exact question.
30. The final USER turn should support natural closure and must not introduce a new
    answer-relevant fact.
31. Do not add disfluency spellings solely to represent paralinguistic style. Audio
    realization controls the style.

SAFETY AND BENIGN CONTENT

32. Keep supporting details benign and ordinary.
33. Do not add self-harm, violence, weapons, illegal activity, credentials,
    sensitive personal data, medical treatment instructions, sexual content,
    targeted political persuasion, or financial transaction instructions.
34. Do not obscure or euphemize unsafe material to bypass safeguards.
35. If a required local fact itself conflicts with these constraints, do not replace
    it with a different fact. The upstream candidate must be regenerated.

ALIGNMENTS

36. Return exact spans for every realized fact.
37. Use zero-based, half-open character offsets.
38. For every span, spoken_text[start_char:end_char] must exactly equal
    surface_text.
39. Every fact_realization span reference must resolve within the same USER turn.
40. Every evidence alignment must use the designated USER turn.
41. Speech-semantic evidence uses semantic_span and at least one span reference.
42. Speaker, paralinguistic, and environmental evidence use audio_turn or
    audio_span. Their acoustic target must not be lexicalized.
43. acoustic_evidence_refs contains the evidence IDs whose audio requirements apply
    to that USER turn.
44. Do not invent an acoustic evidence reference on a non-evidence turn.

OUTPUT CONTRACT

Return exactly one valid JSON object conforming to the supplied output schema.
Use English for every machine-readable string.
Do not return Markdown, comments, analysis, explanations, or chain of thought.
Do not wrap JSON in a code fence.

Before returning, silently verify:
- all assigned USER turns are present once and in order;
- every required fact is realized exactly once;
- every required evidence item is aligned exactly once;
- word-count ranges are respected;
- acoustic targets are not leaked in text;
- no ASSISTANT text is present;
- all spans and references are exact.
\end{verbatim}
\end{promptbox}

\begin{promptbox}[colback=black!5, colframe=white!40!black, title=Dialogue Generation Prompt: Assistant Replies (evidence-blind)]{}
\scriptsize
\begin{verbatim}
[SYSTEM]

You generate one ASSISTANT text reply for one visible conversation prefix.

You are evidence-blind. The input intentionally excludes benchmark facts, evidence
labels, the gold answer, the final query, acoustic labels, future USER turns, and
other sessions.

ONE JOB, ONE REPLY

1. Generate exactly one ASSISTANT reply.
2. Follow the supplied reply_mode and abstract reply_goal.
3. Use only the visible conversation prefix.
4. Do not generate a future USER turn.
5. Do not mention that information is hidden or that this is a benchmark.
6. The runner may batch many jobs with the same assistant position, but your
   response represents only the current job.

REPLY BEHAVIOR

7. Answer an explicit user question when one is present.
8. Otherwise respond briefly and helpfully in a way consistent with the reply
   contract.
9. acknowledge:
   - acknowledge the user's topic without summarizing the specific details.
10. follow_up:
   - ask one broad, natural question that allows several plausible continuations;
   - avoid a narrow question that demands one specific unseen answer.
11. clarify:
   - ask one concise clarification that is necessary for a helpful response.
12. inform_or_help:
   - offer general information, practical help, or a next step;
   - do not invent personal facts about the user.
13. close:
   - close naturally without recapping user-specific details.

NON-ECHO POLICY

14. Do not unnecessarily repeat names, numbers, dates, times, choices, locations,
    quantities, or other specific personal details from the immediately preceding
    USER turn.
15. Do not summarize the user's statement.
16. Do not confirm a personal detail by restating it.
17. Reuse a detail only when it is strictly necessary to answer an explicit
    question, and use the minimum wording required.
18. Do not introduce a competing or contradictory personal fact.

ACOUSTIC BLINDNESS

19. Do not comment on the user's voice, identity, emotion, confidence, hesitation,
    speaking rate, vocal effort, accent, or delivery.
20. Do not mention background sounds, acoustic scenes, recording quality, or
    inferred physical location.
21. Do not guess who the speaker is.

STYLE

22. Write natural English suitable for a helpful voice assistant.
23. Keep within the supplied word-count range.
24. Do not use headings, bullet lists, stage directions, citations, or quotation
    marks around the reply.
25. Do not explain your reasoning.
26. Do not add safety warnings unless the visible user content genuinely requires
    them.

OUTPUT CONTRACT

Return exactly one valid JSON object conforming to the supplied output schema.
Copy job and turn identifiers exactly.
Use English for every machine-readable string.
Do not return Markdown, comments, analysis, or chain of thought.
Do not wrap JSON in a code fence.

Before returning, silently verify:
- the reply matches the visible prefix;
- the reply matches reply_mode and reply_goal;
- no hidden information is assumed;
- no unnecessary personal-detail echo occurs;
- no acoustic commentary occurs;
- identifiers and word_count are correct.
\end{verbatim}
\end{promptbox}

\subsection{Quality-Control Prompts}
\label{app:prompt_qc}

The three question-family gates of \S\ref{sec:qc} run on the assembled history
and are reproduced here.
The answer-validity gate receives the full transcript and no audio; the
acoustic-necessity gate receives the primary audio and its paired control; the
full-history gate receives the complete transcript and every user-turn
waveform of the 64K history. Because histories are nested, the 64K history
contains every session of the shorter ones, and the gate reports the earliest
budget at which a problem appears.

\begin{promptbox}[colback=black!5, colframe=white!40!black, title=Quality-Control Gate: Answer and Operation Validity]{}
\scriptsize
\begin{verbatim}
You are a quality-control judge for a benchmark of long-term spoken conversational
memory.

The question logic, evidence construction, and individual sessions have already
passed earlier quality-control stages.

Your task is only to verify that the **final benchmark item remains semantically
valid after history construction**.

Use only the supplied benchmark materials. Do not use outside knowledge to fill
missing evidence.

You will receive:

* the question;
* the intended gold answer;
* the designated evidence;
* the final assembled context or relevant QC views;
* optionally, a paired answer-refusal (AR) item.

Evaluate the following.

1. **Final answerability**
   With the designated evidence available, does the final answerable item support
   one uniquely justified gold answer?

2. **Evidence necessity**
   Without the designated answer-critical evidence, is the same gold no longer
   uniquely supported?

3. **No semantic shortcut**
   Can the gold be obtained from non-evidence context or another unintended route?

4. **Paired AR validity**
   If an AR item is supplied, is the remaining context genuinely insufficient to
   justify the original gold, without supporting another unique answer?

   This gate sees text only, with no audio. When the sidecar reports
   `ar_delta_channel` as `audio_only` (derivation methods `cue_neutralization` and
   `binding_removal`), the AR delta is a re-rendered WAV: the turn, its words and
   the QC transcript are unchanged by design, and the cue markup in the transcript
   describes the recipe rather than the delivered audio. You therefore cannot
   observe that delta here. Return `"not_applicable"` for this check, unless the
   remaining **text alone** pins the gold down, which is a real defect and a
   `"fail"`. [...]

   When `ar_delta_channel` is `text_and_audio` (`partial_evidence_removal`), the
   removed sessions really are gone from the supplied history and this check
   applies in full.

For each applicable check, return:

* `"pass"` if the condition is clearly satisfied;
* `"fail"` if it is clearly violated;
* `"indeterminate"` if the supplied material is insufficient for a reliable
  judgment.

Return JSON only:

{
  "family_ref": "...",
  "checks": {
    "final_answerable": "pass|fail|indeterminate",
    "designated_evidence_necessary": "pass|fail|indeterminate",
    "no_semantic_shortcut": "pass|fail|indeterminate",
    "paired_ar_valid": "pass|fail|indeterminate|not_applicable"
  },
  "derived_answer": "... or null",
  "brief_reason": "...",
  "status": "pass|fail|indeterminate"
}

`status` is `"pass"` only when all applicable checks pass.
\end{verbatim}
\end{promptbox}

\begin{promptbox}[colback=black!5, colframe=white!40!black, title=Quality-Control Gate: Acoustic Necessity]{}
\scriptsize
\begin{verbatim}
You are a quality-control judge for a benchmark of long-term spoken conversational
memory.

The audio assets and their individual acoustic properties have already passed
earlier audio quality control.

Your task is only to determine whether the **target acoustic dimension is genuinely
necessary for answering the final benchmark question**.

You will receive:

* the question and intended answer;
* the target evidence type;
* the validated primary audio condition;
* the validated control or ablated audio condition;
* the relevant transcript or semantic context.

Evaluate the following.

1. **Primary supports the answer**
   Does the primary audio condition contain the target acoustic information needed
   to support the intended answer?

2. **Control isolates the target dimension**
   Does the control remove or change the intended acoustic evidence while preserving
   the non-target information needed for a fair comparison?

3. **Ablation breaks answerability**
   After the target acoustic information is removed or changed, is the original
   answer no longer uniquely supported?

Use the following interpretation:

* `speaker`: compare the original speaker binding with the speaker-swapped control.
* `paralinguistic`: compare the original vocal cue with the cue-neutralized control.
* `environmental`: compare the original environmental audio with the
  environment-removed control.
* `speech_semantics`: use only when a semantic clue-removal test is explicitly
  requested.

For Speaker, Paralinguistic, and Environmental items, fail if the transcript alone
already uniquely determines the answer.

Do not re-evaluate general audio quality, speaker consistency, or cue
recognizability unless they directly prevent this task-level comparison.

Return JSON only:

{
  "family_ref": "...",
  "evidence_type": "speech_semantics|speaker|paralinguistic|environmental",
  "checks": {
    "primary_supports_answer": "pass|fail|indeterminate",
    "control_is_valid_ablation": "pass|fail|indeterminate",
    "ablation_breaks_answerability": "pass|fail|indeterminate",
    "transcript_only_insufficient": "pass|fail|indeterminate|not_applicable"
  },
  "brief_reason": "...",
  "status": "pass|fail|indeterminate"
}

`status` is `"pass"` only when all applicable checks pass.
\end{verbatim}
\end{promptbox}

\begin{promptbox}[colback=black!5, colframe=white!40!black, title=Quality-Control Gate: Full-History Validity]{}
\scriptsize
\begin{verbatim}
You are a quality-control judge for a benchmark of long-term spoken conversational
memory.

Individual questions, evidence sessions, haystack sessions, filler sessions, and
audio assets have already passed earlier quality-control stages.

Your task is only to determine whether **new validity problems emerge after all
components are assembled into the complete long conversation history**.

You will receive:

* the question and intended gold answer;
* the complete assembled 64K history, with user turns represented by QC-only
  transcripts and assistant turns represented by their actual text;
* a minimal QC sidecar identifying designated evidence, the AR delta, and tier
  introductions. These annotations are for judging only and are not part of the
  conversation history;
* timestamps and an `introduced_at_tier` value for each session;
* optionally, a paired answer-refusal (AR) history.

Do not infer evidence or filler labels from text embedded in the history. Do
not treat the QC sidecar as benchmark model input. Audio attachments, when
present, are the actual user-turn WAVs corresponding to the history; an
attachment carrying a `turn_ref` belongs to exactly that turn.

Evaluate the following.

1. **Non-evidence leakage**
   Does any haystack or filler content directly or indirectly reveal the gold answer
   or required evidence?

2. **Joint shortcut**
   Can multiple non-evidence sessions, when combined, reconstruct the answer even
   though they are individually harmless?

3. **Assistant echo**
   Does an assistant response expose answer-critical information that was intended
   to remain only in user speech or audio?

4. **Answer-relevant contradiction**
   Does the assembled history introduce a persona, factual, or temporal
   contradiction that changes or destabilizes the intended answer?

5. **Alternative answer route**
   Does the full history provide another legitimate route to the gold without using
   the intended answer-critical evidence?

6. **AR residual inference**
   If a paired AR history is supplied, can the answer still be reconstructed from
   what remains? The sidecar states `ar_derivation_method`, and the three methods
   change what "remains" means:

   * `partial_evidence_removal` -- whole sessions are deleted. They are listed in
     `ar_removed_session_refs` and are genuinely absent from the history.
   * `cue_neutralization` -- the answer-critical turn **stays**, with the same
     speaker and the same words, and only its audio is re-rendered with the
     acoustic cue removed. Listed in `ar_replaced_turn_refs`.
   * `binding_removal` -- the answer-critical turn **stays**, with the same words,
     and only the voice is swapped to a different speaker. Listed in
     `ar_replaced_turn_refs`.

   For the two replacement methods, [...] judge the delivered audio: the active
   attachment for that `turn_ref` is the AR rendition,
   and an `ar_source_contrast` attachment for the same `turn_ref` is the
   pre-replacement rendition, supplied so you can compare them.

   Fail only when the answer survives anyway: the cue or voice is still audible in
   the AR rendition, another retained session supplies the same acoustic condition,
   or the remaining text alone pins the answer down.

7. **Acoustic-role collision**
   For Speaker, Paralinguistic, or Environmental items, does another non-target
   session accidentally satisfy the same answer-critical acoustic condition and
   create ambiguity or a shortcut?

If a problem first appears only after a longer context tier introduces additional
sessions, report the earliest affected tier.

Return JSON only:

{
  "family_ref": "...",
  "checks": {
    "no_non_evidence_leakage": "pass|fail|indeterminate",
    "no_joint_shortcut": "pass|fail|indeterminate",
    "no_assistant_echo": "pass|fail|indeterminate",
    "no_answer_relevant_contradiction": "pass|fail|indeterminate",
    "no_alternative_answer_route": "pass|fail|indeterminate",
    "ar_no_residual_inference": "pass|fail|indeterminate|not_applicable",
    "no_acoustic_role_collision": "pass|fail|indeterminate|not_applicable"
  },
  "earliest_affected_tier": "8k|16k|32k|64k|null",
  "offending_session_refs": [],
  "brief_reason": "...",
  "status": "pass|fail|indeterminate"
}

`status` is `"pass"` only when all applicable checks pass.
\end{verbatim}
\end{promptbox}

\subsection{Evaluation and Scoring Prompts}
\label{app:prompt_eval}

Every evaluated model receives the same candidate instruction
(Appendix~\ref{app:input}). The evaluation runs of Section~4 use the variant
that permits abstention. The acoustic-evidence validation of
\S\ref{sec:acoustic-validation}, which contains only answerable questions, uses the
variant without abstention.

\begin{promptbox}[colback=black!5, colframe=white!40!black, title=Candidate Instruction (abstention permitted)]{}
\scriptsize
\begin{verbatim}
You are continuing an ongoing conversation that took place across multiple
timestamped sessions.

User turns are provided as audio, and assistant turns are provided as text. The
final user audio is the current query.

Use the complete conversation history to answer the final query. Consider any
relevant information available in the audio itself, not only the words that were
spoken. Different user audio turns may or may not be spoken by the same person.

Base your answer only on the provided conversation. If the conversation does not
contain enough information to determine one unique answer, respond exactly with:

insufficient evidence

Otherwise, provide a concise answer without explaining your reasoning.
\end{verbatim}
\end{promptbox}

\begin{promptbox}[colback=black!5, colframe=white!40!black, title=Candidate Instruction (no abstention)]{}
\scriptsize
\begin{verbatim}
You are continuing an ongoing conversation that took place across multiple
timestamped sessions.

User turns are provided as audio, and assistant turns are provided as text. The
final user audio is the current query.

Use the complete conversation history to answer the final query. Consider any
relevant information available in the audio itself, not only the words that were
spoken. Different user audio turns may or may not be spoken by the same person.

Base your answer only on the provided conversation. Always give your best answer,
even if you are not certain. Provide a concise answer without explaining your
reasoning.
\end{verbatim}
\end{promptbox}

\begin{promptbox}[colback=black!5, colframe=white!40!black, title=LLM-as-Judge Prompt (answerable items)]{}
\scriptsize
\begin{verbatim}
You are evaluating the correctness of a model's answer to a conversational memory
question.

You are given:
- the final question,
- the reference answer,
- the expected answer type,
- and the model's response.

Determine whether the model's final committed answer is semantically equivalent to
the reference answer.

Evaluation rules:

1. Judge correctness by meaning, not exact wording.
2. Accept harmless differences in capitalization, punctuation, articles, number
   formatting, or equivalent wording.
3. For numerical answers, accept numerically equivalent expressions.
4. For unordered sets, all required elements must be present and no incorrect
   elements may be added.
5. For ordered answers, the required elements and their order must both be correct.
6. A response is incorrect if it omits an answer-critical part of the reference
   answer.
7. A response is incorrect if it gives multiple incompatible answers or contradicts
   its own final answer.
8. A response is incorrect if it abstains, says the evidence is insufficient, or
   refuses to answer instead of providing the answer.
9. Additional explanation is allowed only if it does not change, weaken, or
   contradict the final answer.
10. Do not require the model to match the reference answer's exact surface form
    unless the answer specification requires exact wording.

When an ANSWER OPTIONS list is given, these rules apply in addition:

11. The response is correct only if it commits to exactly one option and that option
    is the reference answer.
12. A response that names two or more options, or hedges between them, is incorrect
    even when one of them is the reference answer.
13. A response that does not quote an option verbatim is resolved by meaning: if it
    is semantically equivalent to exactly one option, treat it as that option; if it
    matches none of them, or more than one, it is incorrect.
14. For an ordered option, every element and their order must match; a response that
    names the right sounds in the wrong order is incorrect.

Return only:

{"verdict":"correct"}

or

{"verdict":"incorrect"}
\end{verbatim}
\end{promptbox}

\begin{promptbox}[colback=black!5, colframe=white!40!black, title=LLM-as-Judge Prompt (AR items)]{}
\scriptsize
\begin{verbatim}
You are evaluating whether a model correctly abstained from answering an
unanswerable conversational memory question.

The correct behavior is to indicate that the provided conversation does not contain
enough evidence to determine one unique answer.

Evaluate the model response using the following rules:

1. A response is correct if it clearly states that the answer cannot be determined
   because the available conversational evidence is insufficient, missing,
   incomplete, or ambiguous.
2. The exact phrase "insufficient evidence" is not required. Semantically equivalent
   evidence-based abstentions are acceptable.
3. A generic statement such as "I don't know" is not sufficient unless it clearly
   attributes the uncertainty to missing or insufficient information in the provided
   conversation.
4. A safety, policy, capability, or willingness-based refusal does not count as a
   correct abstention.
5. A response is incorrect if it provides or commits to a specific answer.
6. A response is incorrect if it first states that the evidence is insufficient but
   then guesses or proposes a likely answer.
7. A response is incorrect if it presents several candidate answers instead of
   abstaining.
8. Do not judge whether the question itself is actually answerable. Assume the
   benchmark annotation is correct and evaluate only whether the model response
   exhibits the required evidence-based abstention behavior.

Return only:

{"verdict":"correct"}

or

{"verdict":"incorrect"}
\end{verbatim}
\end{promptbox}

\subsection{Error-Attribution Prompt}
\label{app:prompt_attr}

The auditor is never shown the error taxonomy. It reports where the committed
answer came from, and the class is derived from its record by the fixed rules
described in Appendix~\ref{app:err_procedure}.

\begin{promptbox}[colback=black!5, colframe=white!40!black, title=Error-Attribution Prompt]{}
\scriptsize
\begin{verbatim}
You are auditing one wrong answer given by a speech model to a question about a \
long spoken conversation history.

The conversation is given as sessions in time order. Each session is marked:

  [EVIDENCE]    it carries information the question is about
  [DISTRACTOR]  a similar session that does not carry the answer
  [FILLER]      unrelated small talk, summarised to one line

User turns are verbatim transcripts of what was spoken. Three things about HOW \
a turn sounded are written into the transcript, because the model heard them \
and a reader cannot:

  <|emotion:X|> <|prosody:X|> <|style:X|>   delivery of the text that follows
  [background sound in this session: X]     what could be heard behind the speaker
  Speaker 1 / Speaker 2 / ...               distinct voices, numbered by first
  appearance

The response is already known to be wrong. Do NOT re-judge correctness. \
Determine only WHERE its answer came from.

Return one JSON object and nothing else:

{
  "commits": "specific" | "generic" | "off_type" | "none" | "truncated",
  "answer": "<what the response commits to, in a few words; null if nothing>",
  "evidence_recovered": true | false,
  "source": "evidence" | "other_session" | "absent" | "none",
  "source_session": <the session number the answer came from, or null>,
  "relation": "matches_gold" | "wrong_value_same_session" |
  "same_value_other_session"
              | "earlier_state" | "later_state" | "other_speaker" | "count_off" |
              "unrelated",
  "why": "<one sentence; quote the line the answer came from, if there is one>"
}

Field guidance:

- "commits":
    "specific"  it names a concrete answer of the kind the question asks for.
    "generic"   it only describes the answer in default terms that would fit any
                session -- for a delivery question, "clear", "neutral", "steady",
                "calm and composed"; for a sound question, "some background noise".
    "off_type"  it answers a different question than the one asked: naming the
                topic of the conversation when asked what could be HEARD, or
                comparing what two threads are about when asked to compare their
                sounds.
    "none"      it commits to nothing -- including a bare acknowledgement
                ("I have noted that down"), and including a flat denial that
                anything was heard when an [EVIDENCE] session carries a cue.
    "truncated" the text stops mid-sentence, or is visible reasoning that never
                reaches an answer ("Wait, look at the timestamps: ..."). Use this
                whenever no final answer was produced because the text ran out.

- "evidence_recovered": did the response show it identified the answer-critical
  cue -- the right sound, the right delivery, the right speaker -- even if the
  final answer is wrong, incomplete or missing? This is separate from "commits":
  a response can name the right sound and still fail to give the count asked for.

- "source": "evidence" if the committed answer appears in a session marked
  [EVIDENCE]; "other_session" if it appears in a [DISTRACTOR] or [FILLER]
  session; "absent" if it appears nowhere in the conversation -- including when
  the response invents an inventory of sounds or speakers that no session has;
  "none" if nothing was committed.

- "relation" is with respect to the REFERENCE ANSWER. Use "earlier_state" or
  "later_state" only when the committed answer is a real value of the same thing
  the question asks about, at a different point in time. Use "count_off" when
  the answer is a number and the members counted are real.

Two worked examples:

  Question asks which route was chosen in the session with water drops; reference
  "market route"; response "canal route"; Session 3 [DISTRACTOR] says "for the
  gallery stop, maybe we ought to choose the canal route", Session 66 [EVIDENCE]
  carries the water drops and says "market route".
  -> {"commits":"specific","answer":"canal route","evidence_recovered":false,
      "source":"other_session","source_session":3,
      "relation":"same_value_other_session","why":"..."}

  [...]
\end{verbatim}
\end{promptbox}

\end{document}